# Firstborn Advantage in the Ivory Tower: Mass Science, Expanding Scholarly Families, and the Reshaping of Academic Stratification

Likun Cao[*‡], Jie Hua[†], James Evans[‡§1]

[*]Purdue University, [†]Renmin University, [‡]University of Chicago, [§]Santa Fe Institute

[1]Correspondence to James Evans, Knowledge Lab and Sociology, University of Chicago at jevans@uchicago.edu.
Likun Cao and Jie Hua contributed equally to this project.
Acknowledgement: We sincerely thank Yuanyi Zhen and Yuting Chen for their constructive comments and insightful feedback during the preparation of this manuscript. The author bears sole responsibility for the content.

# Firstborn Advantage in the Ivory Tower: Mass Science, Expanding Scholarly Families, and the Reshaping of Academic Stratification

## Abstract

This paper investigates the mechanisms underlying scientific stratification in the era of transition from elite to mass science. Existing scholarship has largely examined scientific stratification through the Matthew effect framework at the individual, institutional, and lineage levels, but this theoretical lens has grown limited in today's academic landscape, where mass, team-based, and lab-centered research has become the dominant mode of knowledge production. As scientists increasingly share institutional and lineage backgrounds, considerable variation within these units remains unexplained. We propose a new framework that integrates concepts and methodological tools from demography into the social study of science. Drawing on the parallel between biological families and scholarly lineages as fundamental units of reproduction, we adapt the concept of birth order to examine how the sequence of doctoral students within a lineage shapes their career trajectories. Using data on more than one million U.S. doctoral graduates, our analysis shows that, much like in biological families, later students systematically perform worse than earlier ones across multiple dimensions of academic achievement, both short and long term. Examining the underlying mechanisms, we find that later students receive less cognitive stimulation from mature scholars and instead more from peers, and specialize in narrower intellectual domains as senior siblings occupy adjacent territories. By the time they enter a lineage, the networks and resources of the lineage are more likely to have plateaued. These factors constrain their intellectual development as independent scholars. By introducing a demographic framework into the study of science, this paper offers a new perspective on scientific stratification and demonstrates how demographic concepts can be fruitfully extended to analyze broader social and epistemic systems.

**Keywords:** scientific stratification; birth order; knowledge production; science of science; demographic process

Mechanisms of stratification within the scientific profession have been a central topic in the social study of science since Merton (1979). Status processes determine not only who is rewarded within the academic system, but also who gains the resources and authority to reproduce or reshape its structure. Merton largely described the stratification system of science as driven by Matthew effects: scientists who are already reputable, or with a reputable affiliation, will also attract more resources, citations, and attention in the future (Merton 1979). More recent empirical studies, whether from the sociology literature or from large-scale computational analyses in the science of science, largely build on this perspective and confirm the existence of Matthew effects at multiple levels. Students from prestigious institutions are more likely to secure positions at similarly prestigious institutions (Keith and Babchuk 1998; Clauset et al. 2015; Cole 1973), and protégés from reputable lineages benefit not only from technical skills and tacit knowledge but also from the social networks of their advisors (Wang et al. 2021; Lovitts 2002; Chariker et al. 2017; Barclay and Kolk 2015).

While insightful, this perspective faces growing theoretical limitations in the changing landscape of science. First, stratification among scientists unfolds across their entire careers, not only upon hiring. Given that 80% of domestically trained faculty come from only 20.4% of institutions (Pampel and Hunter 2012), institutional background or lineage alone may not be a sufficiently strong predictor of later stratification dynamics, as the pool of surviving scientists shows limited variation in their backgrounds.

This limitation becomes more salient as science shifts from an elite, mentorship-based model to a mass, institutionalized system (Geiger 2004; Etzkowitz et al. 2000). With changes to the distribution of resources in higher education, as well as the rise of team and collaborative science, prestigious institutions and lineages now produce disproportionately large numbers of protégés, pushing students from non-elite schools out of the system (Malmgren et al. 2010; Long and McGinnis 1981; Zhang et al. 2022). Consequently, among surviving individuals, variation that once existed primarily between institutions and lineages is now increasingly found within them, a trend expected to intensify in the foreseeable future.

Under these circumstances, while the classic structural framework of scientific stratification remains useful for empirical exploration, its explanatory power for understanding scientific stratification may be gradually weakened. In this paper, we propose an alternative theoretical approach: to conceptualize science as a population process, and discuss the uneven distribution of life chances among scientists through a demographic lens. We illustrate this approach through an empirical exploration of scientific reproduction, where we introduce the demographic concept of birth order, i.e., the chronological sequence of children within a family, to examine how students within the same institution and lineage could systematically receive unevenly distributed career opportunities. Our results reveal striking parallels between biological and intellectual families. Within the same lineage, early and later students show systematic differences in career outcomes, with early students consistently achieving better performance across multiple dimensions of professional success.

Why is this the case? To explore the underlying mechanisms, we draw on four of the most influential explanations of birth order effects from the demographic literature: the resource dilution theory, the confluence model, the role specialization perspective, and the family dynamics perspective. To test these hypotheses, we construct indirect proxies based on students' thesis records and publication data. These variables partly reflect the support and guidance students receive from their mentors and peers, as well as the intellectual identities formed during their graduate training. Although they serve only as indirect measures of students' experiences in graduate school, they nonetheless provide valuable insight into the mechanisms of stratification that unfold within their training experiences.

Our empirical analyses yield differentiated support across the four mechanisms: strongest for role specialization, partial for both resource dilution and the confluence model, and modest for family dynamics. Although later students receive comparable levels of resource investment and cognitive stimulation from their advisors, they typically join larger cohorts, collaborate more frequently with intellectual siblings, and face more intensive competition for resources and attention. This competitive pressure pushes them to differentiate by carving out narrower and more specialized niches within the lineage's intellectual territory. When they join the lineage, they are also more likely to enter an environment marked by slowing expansion but growing managerial and quality-control demands, conditions that possibly constrain their autonomy and intellectual range.

By applying a theoretical framework from demography to the social study of science, this study contributes to both domains at the same time. It illuminates both the parallels and distinctions between population and scientific systems, extends the use of demographic theories in understanding scientific communities, and introduces a new dimension of stratification for analyzing scientific careers. In addition, we engage more broadly with the stratification literature, drawing on insights from the recent emerging relational inequality theory (Tomaskovic-Devey and Avent-Holt 2019; Avent-Holt and Tomaskovic-Devey 2019). By applying this perspective to the stratification of science, we show that inequality is not merely the result of structural resource allocation among individuals but is actively produced and reproduced through social interactions within local organizational contexts, such as everyday exchanges and communications between mentors and mentees, students and peers.

In what follows, we first synthesize the literatures on academic stratification and demographic theory. We then describe our data and within-lineage fixed-effects design, followed by the main effects and mechanism tests. We close by drawing implications for the sociology of science and for innovation policy.

## Academic Stratification and Career Outcomes in Science

Stratification in scientific careers has long been a central topic of scholarly discussions in the sociology of science. The mechanisms of this stratification largely determine who gains access to resources, recognition, opportunities, and academic authorities (Cole 1973). As in other employment settings, merit-based evaluations play an important role in academia, with publications and scholarly output forming the foundation of academic hierarchies (Li and Agha 2015). Yet merit-based hiring and promotion alone cannot fully account for career opportunities or the distribution of resources in science. In fact, the variance explained by merit-based standards remains modest (Keith and Babchuk 1998; Clauset et al. 2015; Elder and Kozlowski 2025).

Institutional prestige is one of the most significant factors shaping academic stratification: graduates from prestigious institutions consistently secure better career opportunities. Clauset, Larremore, and Arbesman (2015) find that faculty hiring is highly concentrated, with only 25% of doctoral programs producing 71–86% of tenure-track professors across history, computer science, and business. A similar study of all tenure-track faculty in U.S. Ph.D.-granting universities between 2011 and 2020 found that 80% of domestically trained faculty came from only 20.4% of institutions (Pampel and Hunter 2012). Although institutional prestige and scholarly output are often intertwined, institutional prestige alone has been found to be a stronger predictor of career outcomes than scholarly achievement, both at the individual level of job search and collective level of faculty hiring (Keith and Babchuk 1998; Clauset et al. 2015).

A second source of divergence in career opportunities arises from students' choice of advisor. Advisorship plays a crucial role in shaping scholars' intellectual orientations and achievements. The most direct pathway is through professional training: advisors develop research skills, foster creativity, and enhance the publication output of their protégés (Wang and Shibayama 2022; Li and Fernandez 2024; Ma et al. 2020). These publication advantages can then translate into career rewards.

While advisors are formally assigned to oversee students' academic progress, their influence extends far beyond technical training: mentors also transmit tacit knowledge about the scientific profession (Polanyi and Sen 2009). Through interactions, advisors serve as gatekeepers and socializing agents, introducing students to disciplinary norms, research ethics, professional expectations, and social and networking practices of the field. They also train mentees in supporting skills such as grant writing, conference networking, publication strategies, and subtle social skills, better preparing them for career challenges (Wang et al. 2021; Lovitts 2002; Hofstra et al. 2022). Empirical evidence underscores the importance of their critical role in tacit knowledge transfer. Advisors with more patent filings tend to have students who also file intellectual property (Delgado and Murray 2023), and advisors' networks significantly affect the quality and quantity of social ties their mentees build (Wang et al. 2021). Finally, insights from social network studies apply to the mentor–mentee relationship: ties to prestigious mentors can serve both as channels of resources and as positive signals, influencing outcomes in the job market as well as in broader professional development (Podolny 2001). In these and less tangible ways, advisors can play a pivotal role in shaping their mentees' work and careers: Nobel laureates are more likely to have Nobel laureate ancestors (Chariker et al. 2017), whereas having a lower-status advisor or experiencing poor mentorship can systematically hinder professional development (Sambunjak et al. 2006; Gardner 2007). The influence of advisorship thus resembles parental effects, with prestige and career rewards transmitted across multiple generations (Barclay and Kolk 2015).

## The Growth of Higher Education, the Rise of Collaborative Scientific Culture, and the Expansion of Scholarly Families

The Matthew effect framework theorizes scientific stratification as a process of cumulative advantage: scientists with prior recognition attract disproportionate resources, citations, and visibility, which beget further recognition (Merton 1979). In principle, this logic should operate within institutions and lineages as readily as across them. Yet the framework was developed during the era of elite science, when researchers worked individually or in small groups at elite institutions, and the population of scientists remained small. As a result, empirical studies guided by this framework have predominantly focused on inter-institution and inter-lineage stratification, leaving within-lineage dynamics largely unexamined.

Recent decades have witnessed the emergence of a new dimension of stratification. With the expansion of higher education, growing inequality in scientific resources, and the rise of a collaborative scientific culture, a small number of prestigious institutions and lineages now produce a disproportionately large share of new scholars (Zuckerman 1996; Zhang et al. 2022). As these groups expand, career outcomes have become increasingly differentiated—not only between lineages but also *within the same lineage*, even among students in the same institution and advisory line.

The postwar era witnessed the rapid growth of research universities. Not only did the number of institutions increase, but many made doctoral training a central part of their mission for the first time. Doctoral education and advisorship underwent rapid growth, shifting from an elite, mentorship-based model to a mass, institutionalized system (Geiger 2004; Etzkowitz et al. 2000; Price 1971). In this broad context, two historical trends unfolded and supported the emergence and flourishing of a new institutional ecology.

First, as science became more well-funded and resource-intensive, a small number of elite scientists, often referred to as "stars", increasingly dominated knowledge production, research funding, and doctoral training networks (Zuckerman 1996; Cole 1973). These star scientists not only attracted substantial resources, but also trained disproportionately large numbers of successful protégés (Malmgren et al. 2010; Long and McGinnis 1981; Zhang et al. 2022). As teams expand (Wuchty et al. 2007), a growing share of resource allocation in the scientific system shifts from a market-based mode to a hierarchy-based or network-based mode, reflecting the intertwined dynamics of lab formation and scientists' career cycles (Williamson 1973, 1975). Consequently, much of the variation in career opportunities and outcomes that previously existed between lineages now occurs within them.

Meanwhile, the past few decades have also witnessed the rise of a collaborative culture in science and innovation. The individual-based, elite model of scientific research, which was more common in the humanities and social sciences than in the natural sciences, is increasingly being supplanted by team-based research and "collective intelligence" (Wuchty et al. 2007; Linzhuo et al. 2020; Yang et al. 2024). The lab-centered model, once typical of the natural sciences, has gradually expanded into other disciplines (Becker 2020). In this context, Ph.D. mentees take on dual roles. On the one hand, they are expected to be trained as independent scholars capable of conceiving and leading their own research. On the other hand, they also work as lab researchers, actively participating in collaborative projects, with topics often assigned rather than chosen by themselves. Collaboration between advisors and advisees, as well as among peers, has become more frequent, heightening the importance of team and peers in shaping academic career outcomes (Xing et al. 2025). In this context, the distribution of attention and resources within teams and labs, together with mentees' roles and identities in these settings, can critically influence their intellectual trajectories as a professional scholar.

These trends have produced a new dimension of academic stratification: intra-lineage stratification, which remains relatively underexplored but is becoming increasingly important, especially in the natural sciences where large teams now dominate (Wuchty et al. 2007). While compatible with cumulative-advantage logic, it directs attention to how stratification operates within micro-organizational units that the literature has typically treated as homogeneous.

Given the resemblance between scholarly families and biological families, we draw insights from demography and apply the theoretical lens of family dynamics to the science system. By linking these two theoretical streams, we reframe academic stratification as not only a structural, but also a generational and relational process.

## Birth Order in Scientific Lineages: A Demographic Perspective

In this paper, we adapt the concept of birth order from demography to the social studies of science, examining how resources and identities are distributed among members within the same academic lineage. We take this perspective based on the structural, functional, and social parallels between biological and intellectual families, which both serve as basic units of reproduction within social and knowledge systems, respectively.

Birth order refers to the sequence in which children are born into a family (Sulloway 1999). Scholarly discussions of birth order date back to the late nineteenth century, when firstborns were found to be disproportionately represented among elites (Galton 2018 [1974]). By the mid-twentieth century, with the advent of large-scale censuses and social surveys, a substantial body of research had confirmed the multidimensional effects of birth order. Existing studies show that earlier children typically achieve better outcomes across key social domains, including intellectual development (Zajonc and Markus 1975; Sulloway 2007), educational attainment (Black et al. 2005; Härkönen 2014), health level (Black et al. 2016; Brenøe and Molitor 2018), and socioeconomic status (Behrman and Taubman 1986). The first child particularly outperforms their younger siblings across these dimensions (Ellis 1904; Altus 1965; Bjerkedal et al. 2007; Havari and Savegnago 2022). Although some studies have questioned the causal link between birth order and individual outcomes (Bayer 1966; Rodgers et al. 2000; Bertoni and Brunello 2016), evidence suggests that the difference between early-born children and their siblings is systematic, even when controlling for important confounders and considered only within family (i.e., family fixed effects).

Here we apply the birth order framework to analyze the scientific system based on three justifications. Functionally, both intellectual and biological families serve as fundamental units of reproduction within social systems. Intellectual families transmit scholarly knowledge and academic traditions, just as biological families pass down genes. Structurally, both systems depend on similar social mechanisms for resource allocation and personality development. Just as parents decide how to distribute limited resources and attention among their children, mentors and especially principal investigators (PIs) who direct finite project resources allocate them among their protégés. Finally, both intellectual and biological families exhibit cyclical entry patterns. Children are born sequentially, while students and mentees join advisors in annual cohorts. In both cases, membership and socialization follow similar temporal trajectories. Building on these parallels, we extend the birth order framework to the social study of science. We argue that birth order, along with other demographic concepts, can serve as valuable analytical tools in the emerging era of mass team science, where population dynamics within closely interacting local units increasingly shape the social organization of knowledge production.

Under this framework, we define ***intellectual birth order*** as the chronological sequence in which doctoral students complete their dissertations under the same advisor at the same institution. Analogous to biological siblings, this ordering reflects the relative positions of young scholars within a "sibling" group. When defining a scholarly family, we hold both institution and advisor constant to ensure that the effects we identify are not confounded with institutional and lineage prestige, as these factors have been well documented in prior studies.

Drawing on existing studies of birth order effects, we propose the following hypothesis:

*H1: Earlier students of the same advisor, especially the first student ('firstborn'), will achieve the best career outcomes across multiple dimensions of academic success.*

## Mechanisms Underlying Intellectual Birth Order Effects

Once children gain early advantages, these advantages often accumulate and compound over time (Kidwell 1982; Travis and Kohli 1995; Mechoulan and Wolff 2015). The same pattern applies to students in the scientific system, where early advantages and prestigious academic positions can produce continuous cumulative returns (Merton 1979; Cole 1973; Zhang et al.

2022). But what are the initial forces that lead earlier-born children and students, particularly firstborns, to perform better? Demographic theories offer rich competing explanations. In this paper, we empirically examine the four most important hypotheses: the resource dilution theory, the confluence model, the role-based perspective, and the family dynamics perspective.

***Resource Dilution Theory***

The core premise of the resource dilution theory is that parental resources, such as attention, time, and wealth, are inherently limited (Blake 1981; Downey 2001). Children with more siblings receive fewer resources on average than those with fewer siblings (Jæger 2009). As a result, firstborns enjoy early, and often exclusive, access to familial resources, receiving a concentrated share of parental attention and investment (Öberg 2015; Jensen et al. 2017). This advantage persists even after accounting for family size (Belmont and Marolla 1973; Booth and Kee 2009). The impact of resource dilution is particularly pronounced in larger families, where both sibling number and birth order interact to shape children's developmental trajectories (Bagger et al. 2021).

Given that academic families also face limitations on resources and attention from advisors (i.e., academic "parents"), similar patterns of resource dilution may arise among doctoral siblings. This competition could be even more long-lasting: in some disciplines or countries, early-borns typically remain involved in the distribution of lineage resources after graduation (Chariker et al. 2017). As a result, later students potentially face greater competition. Based on that, we hypothesize that later students will, on average, experience worse career outcomes because they receive less resource investment from the parental generation (H2).

### *The Confluence Model*

The second influential explanation of the birth order effects is the confluence model (Zajonc and Markus 1975; Zajonc and Mullally 1997). The model argues that the intellectual environment of a family changes as new children are born. Firstborns spend more time in an adult-dominated environment, which fosters advanced language and cognitive development, while later-borns grow up in a setting increasingly shaped by interactions with other siblings. Although the confluence model has been challenged for its weak empirical evidence without longitudinal data on parental and child cognitive abilities (Jæger 2009; Gurgand et al. 2023), the model makes an important contribution by identifying the family's early environment as a key mechanism for birth-order effects, particularly in cognitive outcomes.

In academic settings, the confluence model partially overlaps with resource dilution theory. Cognitive stimulation from parents—in this context, advisors—represents a key academic resource. Based on the Confluence Model, we argue that later-born students may receive less cognitive stimulation from their mentor but more cognitive stimulation from peers (H3).

### *Role-based Perspective*

A third explanation for birth order effects comes from an evolutionary psychology perspective. Role-based accounts argue that birth order entails distinctive social and psychological roles, with siblings differentiating themselves to reduce competition for parental attention and resources. Firstborns are often characterized as more responsible and leadership-oriented, serving as role models for younger siblings (Adler 1928; Sulloway 1997). In contrast, later-borns are more likely to seek alternative paths to parental recognition, developing traits such as risk-taking, sociability, or nonconformity (Kidwell 1982; Fukuya et al. 2021).

In academic families, we expect a similar process of role specialization. Although students are typically young adults with relatively stable psychological traits when they join the lineage, they may still develop intellectual identities and pursue distinct research directions that set them apart from their peers. From a role-based perspective, we therefore hypothesize that later students will exhibit greater specialization in their research and develop narrower intellectual foci (H4).

***Family Dynamics Perspective***

A final perspective on birth order effects emphasizes that siblings do not grow up in the same family, even when they share the same household. This is because the family is not a static container but a dynamic system that evolves with the arrival of each child: as parents age, they accumulate parenting experience, change their economic and social circumstances, develop new pedagogical strategies, and modify their interactional styles in response to the children already present (Plomin and Daniels 1987; Dunn 1992). Earlier-born children thus enter a family in one configuration, while later-born children enter a family already transformed by their predecessors; the family environment itself is in motion.

In academic lineages, an analogous dynamic operates with even greater clarity. Each new student enters not a stable lab but one shaped by its research agenda, funding trajectory, and the cumulative impact of all prior students. Earlier students help define the niche; later students inherit and fit in one partially occupied. Whereas role specialization describes how students actively position themselves within the lineage, the family dynamics mechanism describes how the micro environment positions students before any active choice can be made.

We therefore hypothesize that later students will encounter a lab environment that is more settled and more complex: slowing in expansion but rising in managerial and administrative demands (H5). Because the micro-environment of mentorship shapes the questions students pursue, the collaborations they form, and the allocation of their attention, these dynamics carry important implications for mentees' training.

We now turn to a large-scale empirical analysis to test these four competing mechanisms.

## Data and Methods

### *Data*

To examine social processes within intellectual families, we identified all doctoral degree recipients and their advisors, based on theses awarded between 1950 and 2024 by accredited U.S. institutions, as recorded in the ProQuest dataset. This dataset represents one of the most comprehensive collections of Ph.D. dissertations and degrees to date and has been widely used in sociological studies (Heiberger et al. 2021). In this data, each thesis record includes a unique identifier, abstract and full paper, as well as metadata such as the author's name, advisor's name, year of publication, and institutional affiliation. Using advisor names and university affiliations, we linked each student to a distinct intellectual family, and connected them with their "siblings" within the same academic lineage. In total, the dataset includes 1,571,601 doctoral theses submitted between 1950 and 2024. We accessed the data through the University of Chicago's institutional subscription to ProQuest.

We linked both Ph.D. students and their advisors to publication records in OpenAlex (OA), one of the most commonly used datasets used in the science of science studies. Following prior work, we implemented several procedures to improve disambiguation accuracy and minimize potential mismatches. Details of the data curation and matching process are provided in Appendix A. Overall, we identified 321,453 Ph.D. recipients with publication records from OA. The proportion of Ph.D. students with OA publication records is similar to findings reported in previous literature (Heiberger et al. 2021). We further replicate our analyses using Web of Science (WoS) as an alternative data source, with results reported in Appendix F. The two datasets, OA and WOS, differ along a well-known trade-off: WoS offers stricter data curation at the cost of narrower coverage (focusing on journals rather than books and conferences), while OA provides broader coverage with documented quality concerns (Zhou and Sun 2024). Triangulating the two thus offers a meaningful test of whether our findings depend on the choice of bibliometric source.

To reduce noise, we exclude rare instances in which an individual received multiple Ph.D. degrees. Although we account for them when calculating birth order within each intellectual family, we excluded them from regressions, as their trajectories are atypical and their behaviors and choices may differ substantially from those of other scholars. In total, we retained 1,563,872 valid records.

***Explanatory Variables: Intellectual Birth Order***

We defined intellectual birth order by the sequence in which students defended their dissertations, with intellectual "siblings" who share the same advisor and university affiliation. Within the micro-context of intellectual families, the educational environment, institutional resources, and advisor networks are generally more consistent. This operationalization enables us to rule out confounding effects related to institutional and lineage prestige. We illustrate this measure with the upper panel (Panel A) of Figure 1. To reduce data complexity and minimize noise, we grouped intellectual birth order into the following categories: the first student (“firstborns”), students 2–5, 6–10, 11–15, 16–20, 21–30, and those ranked 31 or higher. The distribution of Ph.D. students across these groups is shown in Panel C of Figure 1.

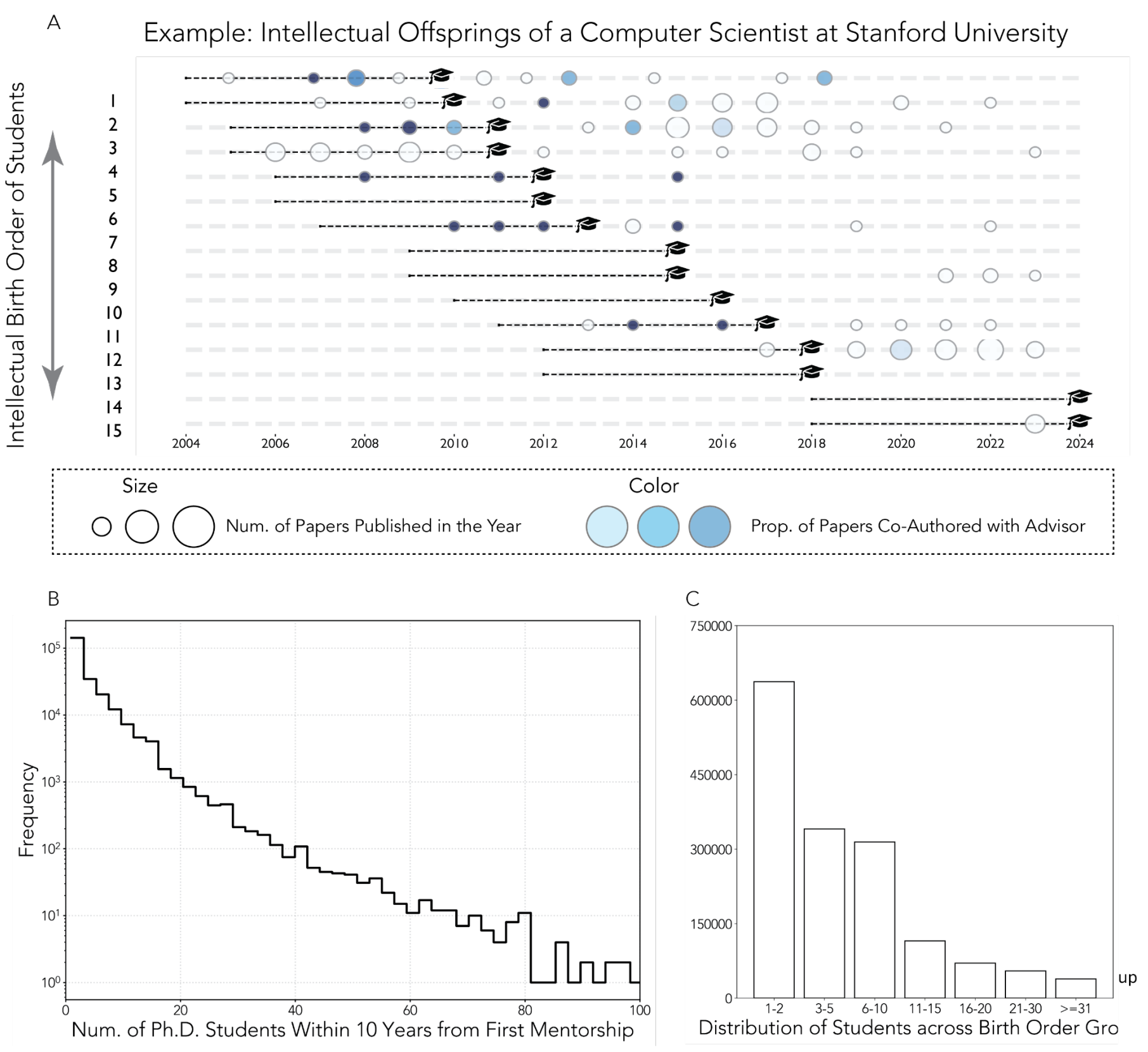

Figure 1. Operationalization of Intellectual Birth Order

Following common practice in demographic research, we include intellectual birth order in our models as dummy variables, using firstborns as the reference group. Two real-world complexities may complicate the identification of birth order as defined here. First, some students may drop out of the Ph.D. program without completing their degree. Second, some individuals may graduate earlier or later than originally planned, thereby slightly altering the birth order within their lineages. To assess the prevalence of these cases and their potential impact on birth order identification, we designed a short survey to capture scientists' recollections of their academic experiences (for details, see Appendix E). The results of the survey show our operationalization can be solid.

***Outcome Variables***

We measure mentees' academic career success using four variables, following prior studies (Heiberger et al. 2021; Xing et al. 2025; Milojević et al. 2018).

***Transition to Advisor (Advisorship).*** The first outcome captures whether a former Ph.D. student has become a primary advisor to a new generation of doctoral students—a key milestone indicating gatekeeping status in the field (Heiberger et al. 2021). To measure this transition, we match the names of former Ph.D. students to those listed as primary advisors in our dataset. In total, we identify 89,572 individuals who have successfully made this transition. Individuals who have not completed this transition by the data collection cutoff (December 2024) were treated as right-censored in survival models.

Ph.D. holders who leave the U.S. do not appear as advisors in our data (Heiberger et al. 2021). Given the central role of the U.S. in global higher education, however, securing a faculty position in the U.S. is generally considered a prestigious milestone, and excluding such cases yields more conservative coefficient estimates rather than more extreme ones.

***Number of Students after Graduation (Students Count).*** Next, we employed an alternative measure of advisorship: the total number of Ph.D. students a scholar supervised after graduation (Xing et al. 2025). This measure offers a continuous indicator of the advisorship activities.

***Number of Publications Within 0–5 Years After Graduation (Short-Term Pubs).*** Ph.D. students are generally expected to establish their careers after graduation, either by independently developing new research projects or by joining another team as postdoctoral researchers. Whether they can successfully achieve this goal serves as a key indicator of their potential to survive and begin a scientific career. To capture early-career productivity, we calculate the number of publications within the first five years after graduation. The number of papers published during this period reflects how quickly individuals transition into their new positions and thrive there.

***Number of Publications Within 5–10 Years After Graduation (Long-Term Pubs).*** Finally, we complement the third dependent variable with a long-term measure: the total number of publications produced 5–10 years after graduation. For the third and fourth variables, we include only individuals who graduated before 2014 to allow sufficient time for observing post-graduation publications. We also applied log transformation to reduce over-dispersion.

***Controls***

We control for several key covariates when predicting scientific career outcomes: (1) mentees' gender, predicted using Genderize; (2) mentees' race, predicted using NamePrism. Both Genderize and NamePrism are widely used tools that infer gender and race based on the statistical distribution of names with global coverage and have been validated in previous studies (Ye and Skiena 2019; Ye et al. 2017). (3) Discipline of dissertation, mapped using the UCSD map of science (Börner et al. 2012); (4) mentees' prior scientific training, measured by number of publications before their Ph.D. years; and (5) advisor's mentoring experience at the time of the thesis defense, measured as the number of years between their first student's defense and the focal student's defense year. For variables (4) and (5), we applied log transformation to reduce over-dispersion.

Because we use intellectual family fixed effects based on advisor-affiliation pairs, our models absorb both advisor-level characteristics, such as advisors' gender and race, and affiliation-level factors, such as university funding sources and institutional tier, into model specification. We also included graduation year as dummy variables for each year from 1950 to 2024. This approach enables comparisons of scholars' achievements within their own cohort.

### *Variables for Mechanism Analysis*

In the final section of this paper, we examine four mechanisms discussed in existing literature. Given the limited scope of our data, we construct measures from mentors' and students' academic and professional records. While these measures help us probe the social dynamics behind our observations, they should be regarded as imperfect proxies rather than direct indicators of the underlying mechanisms we specified earlier.

***Resource Dilution Theory.*** While we cannot directly observe how each scholarly family allocates its resources, the publication structure offers partial insight into the extent to which focal students benefit from their mentors' support and investment. We measure this through two indicators: (1) the number of papers co-authored with the mentor on which the student is first or corresponding author, and (2) the average impact factor (i.e., 2-year citedness) of the venues in which these co-authored papers appear. Both measures are calculated solely based on papers published during graduate school. Publishing in prestigious venues often requires substantial support from mentors, including funding, data access, equipment, and dedicated mentorship time. Serving in key authorial roles on coauthored papers also reflects the mentor's willingness to invest in and advance the student's career development.

In a series of robustness tests, we further standardized these measures by converting them into *z*-scores within each sibship, and re-estimated our models (See Appendix C). The results remain consistent with those reported in the main text.

***The Confluence Model.*** The Confluence model suggests that early-born children receive more cognitive stimulation from their parents, while later-born children are more likely to grow up in environments where sibling interactions play a larger role. In this context, we measure: (1) number of publications during the Ph.D. years coauthored with advisors but not siblings; (2) number of papers coauthored with intellectual siblings but not advisors; and (3) proportion coauthored with both advisors and siblings. We expect early-born students to have higher values for (1), while later-born students are likely to have higher values for (2) and (3). In Appendix C, we replicate our models with proportions, rather than absolute number of papers, as alternative measures. The results remain consistent with those reported in the main text.

***Role Specialization.*** The role specialization explanation claims that siblings differentiate themselves to reduce competition for parental attention and resources (Hotz and Pantano 2015). While role differentiation in biological families often yields personality and psychological trait differences, such as leadership versus rebellion (Fukuya et al. 2021; Kidwell 1982), specialization in intellectual families can be more directly achieved through choice of research directions: students differentiate themselves from their peers by pursuing distinctive topics. To capture these dynamics, we first measure (1) the breadth of knowledge in each student's dissertation. We map the research topics within each dissertation onto the categories of the UCSD Map of Science (Börner et al. 2012), which is constructed based on empirical distances between knowledge domains, and calculate knowledge breadth as the total number of knowledge fields covered; (2) the similarity between a focal student's dissertation abstract and those of their cohort (within ±2 years) using Jaccard similarity at the token level; and (3) the intellectual distance between the focal student and their advisors, computed as cosine distance between the dissertation abstract and the aggregated abstracts of the advisor's prior publications in semantic space. While the first two measures capture differentiation from intellectual siblings, this third measure captures differentiation from the advisor's established research program. We expect later students to be more specialized in their Ph.D. scholarship, with a narrower scope, and lower similarity to peers.

***Family Dynamics Perspective.*** The family dynamics perspective predicts that later students enter lineages whose social networks, resource flows, and managerial demands have shifted since the lab was founded. We construct two lab-level proxies for this evolving environment, each computed from the advisor's bibliometric record. (1) Research Network Expansion: the change in average team size on the advisor's co-authored papers between the five years preceding and during the student's training, capturing whether the lab's collaboration network is still expanding. (2) Funding Expansion: the change in the advisor's acknowledged grant counts across the same window, capturing whether the lineage is still scaling its resource base. As with our other mechanism variables, these are imperfect bibliometric proxies for the underlying construct of lineage maturation. Consistent with H5, we expect later students to enter lineages with lower scores on the two expansion measures.

### *Analytical Strategy*

Transition to advisorship reflects both who becomes an advisor and how quickly they reach this milestone. For this outcome, we apply an event history analysis using the stratified Cox proportional hazards model (Heiberger et al. 2021; Barclay and Kolk 2015; Blossfeld and Rohwer 2007). This approach allows us to estimate the factors that shape the timing of the transition to advisorship without making assumptions about the baseline hazard function. Formally, the hazard function for individual *i* at time *t* is specified as:

$$h_{i,j,t} = h_{j,t} \exp(\beta_0 + \beta_1 \times Order_{i,j} + \beta_2 \times Control_i + T_i + \epsilon_j) \qquad (1)$$

Here, $h_{i,j,t}$ is the hazard rate for individual *i* from family *j* at elapsed time *t* since *i* enters the risk set, $h_{j,t}$ is the baseline hazard function for family *j*. We define the intellectual family ID as the unique strata, allowing each family to have its own baseline hazard function. Mathematically, this is equivalent to including family fixed effects in the model estimation. Individuals are tracked from the year of their Ph.D. graduation until the year they advise their first Ph.D. student, with right-censoring applied to those who have not made the transition to advisorship by the end of the observation period. This model is estimated using the stcox command in STATA 18.0.

For the other three dependent variables, *Student Num.*, *Short-Term Pubs., and Long-Term Pubs.,* we specify the model as:

$$Y_{i,j} = \beta_0 + \beta_1 \times Order_{i,j} + \beta_2 \times Control_i + T_i + \epsilon_j \qquad (2)$$

where $Y_{i,j}$ refers to the dependent variables, $Order_{i,j}$ represents the intellectual birth order of student *i* from intellectual family *j*, $Control_i$ is a vector of covariates, $T_i$ is the fixed effect for graduation year of student *i*, and $\epsilon_j$ denotes the fixed effect component for intellectual family, and thus sibling group, *j*. The three dependent variables in this specification are continuous. Accordingly, we use ordinary least squares (OLS) regression for estimation. The first set of models was estimated using the reghdfe package in STATA 18.0, which allows for the inclusion of high-dimensional fixed effects.

In both sets of models, standard errors are clustered at the family level (Barclay and Kolk 2015; Lillehagen and Isungset 2020).

## Results

### Growth of Higher Education and the Expansion of Intellectual Families: 1950-2024

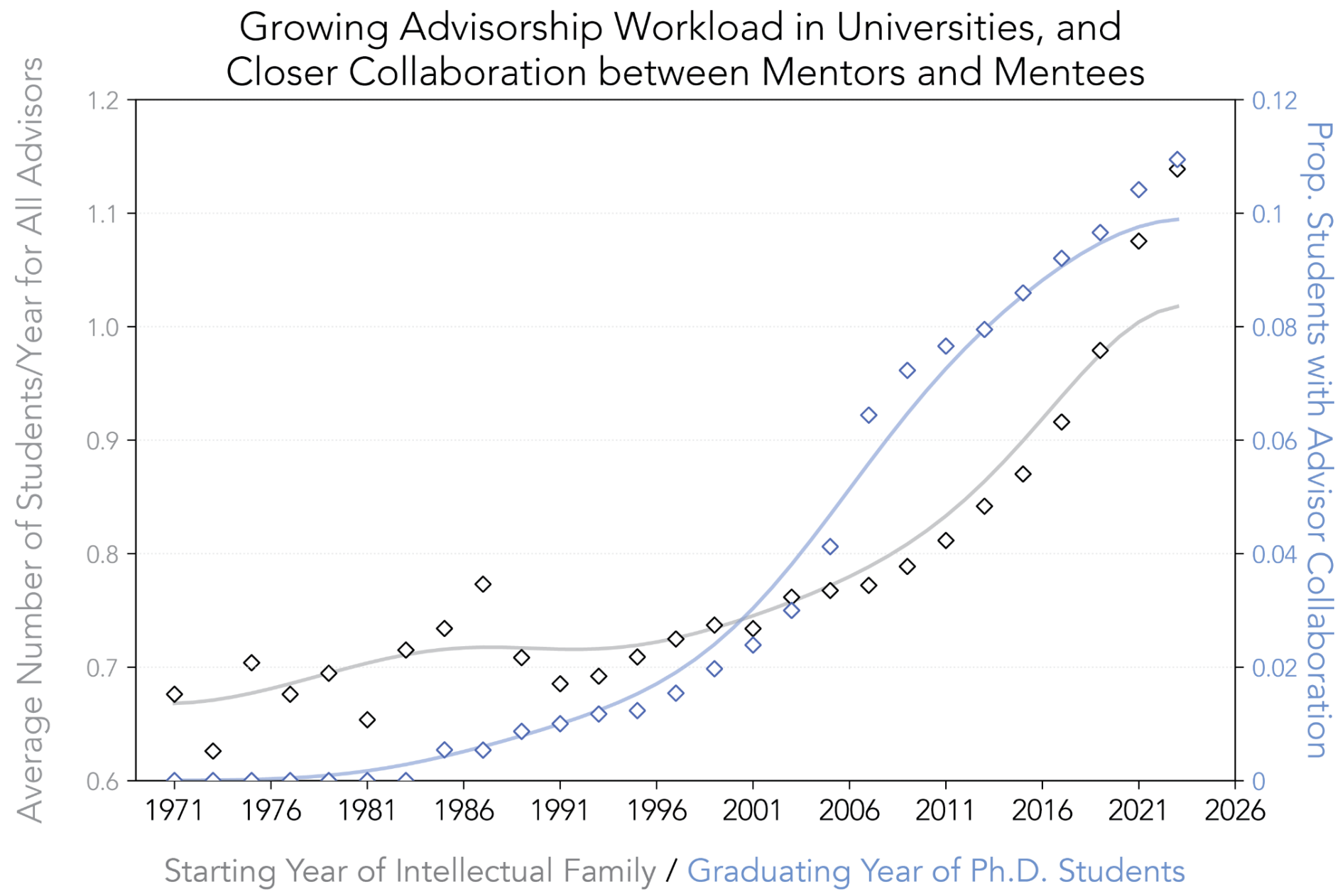

Figure 2. Growing Importance of Advisorship in the U.S. Universities: 1950-2024

Our study is situated within the broader social context of the expansion of Ph.D. programs in the United States, where higher education has become increasingly institutionalized and professionalized. The postwar era saw substantial growth in research universities and the integration of doctoral training into the core mission of academic knowledge production (Geiger 2004). This transformation has fundamentally reshaped the nature of Ph.D. advisorship. As shown in Figure 2, the average advisorship workload (the grey line) steadily increased by approximately 40% from 1971 to 2024. Scholars who began their careers more recently tend to advise more doctoral students per year.

The growth of Ph.D. programs reflects not only a quantitative increase in advisorship workload but also a shift in the nature of advisorship, characterized by the rise of lab-centered mentoring models and an expanding culture of scientific collaboration (Becker 2020; Wuchty et al. 2007). In 1971, very few Ph.D. students co-authored papers with their advisors during their doctoral studies. By 2024, this proportion has grown to approximately 11% and continues to rise, as illustrated by the blue line in Figure 2.

This empirical pattern underscores the need to reconsider academic stratification from a new perspective. In traditional mentoring systems, where academic training followed an apprenticeship-like model—each advisor supervising only a few students, with advisorship largely realized through students' independent work—academic stratification has typically been analyzed in terms of differences between institutions and intellectual lineages (Clauset et al. 2015; Way et al. 2019; Ma et al. 2020). With the rise of lab-centered mentoring structures and the increasing concentration of advising responsibilities, however, intra-lineage variation continues to grow, creating a new dimension of stratification over time.

***Career Advantages of Early Disciples***

We begin by presenting the descriptive statistics of the main variables in Table 1. The proportion of individuals actively publishing steadily declines from 15.94% in the first five years after graduation to 14.06% in the five to ten years following graduation. Yet those who remain in academia tend to produce more publications on average over the long term, with the five-year average total publications increasing from 1.533 to 2.343. Notably, the standard deviation in research output also grows over time, rising from 11.072 to 23.932, indicating a widening gap in productivity among those who persist in academic careers. Only a small proportion of Ph.D. graduates (5.7%) successfully transition to becoming Ph.D. advisors, similar to the number recorded in previous literature (Heiberger et al. 2021). This number shows that intellectual reproduction is concentrated primarily at the top of the academic system.

We then examine how intellectual birth order influences career opportunities and academic status within each lineage. Table 2 reports the regression results, consistently showing a negative and statistically significant relationship between intellectual birth order and career outcomes across specifications. Across all four dependent variables, we observe a clear decreasing gradient in the coefficients. Therefore, Hypothesis 1 is supported, indicating a birth order effect in intellectual families, similar to that observed in biological ones.

When including family fixed effects, the differences between firstborns and students ranked 2nd to 5th within each intellectual family is mostly marginally significant, but a consistent disadvantage emerges starting with the 6th student in each sequence. Specifically, when compared with the first student in each intellectual family, students ranked 6th to 10th have a 12.4% lower hazard ratio to become Ph.D. advisors themselves ( $e^{-0.132}$-1 = -12.4%), and, on average, supervise 0.04 fewer students over the course of their careers, compared to early students. They also publish 1% fewer papers in both short term and long term (with $\beta$= -0.009+ and -0.007, therefore $e^{-0.009}$-1= -1%) on average. These disadvantages become more pronounced for students ranked further down in intellectual families. For students with an intellectual birth order ≥ 31, they produce 2.37% fewer papers in the short run, 2.27% fewer papers in the long run, have a 17.96% lower hazard ratio to become advisors themselves, and have 0.165 fewer students during their lifetime.

Table 1. Descriptive Statistics for Ph.D. Graduates in United States

| **Variables** | **Full Sample Size** | **Mean/N** | **SD/%** |
|---|---|---|---|
| *Outcome Variables (MEAN, SD): Short-Term Pubs is calculated for individuals who graduated in 2019 or earlier; Long-Term Pubs is calculated for individuals who graduated in 2014 or earlier.* | | | |
| Transition to Advisor | 1,563,872 | 0.057 | 0.232 |
| Students Count | 1,563,872 | 0.523 | 3.924 |
| Short-Term Pubs | 1,061,378 | 1.533 | 11.072 |
| Log(Short-Term Pubs) | 1,061,378 | 0.280 | 0.750 |
| Long-Term Pubs | 1,061,378 | 2.343 | 23.932 |
| Log(Long-Term Pubs) | 1,061,378 | 0.303 | 0.851 |
| *Intellectual Birth Order (N, %)* | | | |
| 1 | 1,563,872 | 327,800 | 20.96% |
| 2-5 | 1,563,872 | 645,130 | 41.25% |
| 6-10 | 1,563,872 | 313,049 | 20.02% |
| 11-15 | 1,563,872 | 114,808 | 7.34% |
| 16-20 | 1,563,872 | 70,029 | 4.48% |
| 21-30 | 1,563,872 | 54,536 | 3.49% |
| ≥31 | 1,563,872 | 38,520 | 2.46% |
| *Controls: Continuous (MEAN, SD)* | | | |
| Log(Mentor Experience) | 1,563,872 | 1.669 | 1.146 |
| Log(Previous Publications) | 1,563,872 | 0.019 | 0.152 |
| Year | 1,563,872 | 2007.683 | 10.554 |
| *Demographics of Mentees (N, %): Race* | | | |
| White | 1,563,872 | 1,102,921 | 70.53% |
| Asian/Pacific Islander (API) | 1,563,872 | 367,831 | 23.52% |
| Black | 1,563,872 | 16,789 | 1.07% |
| Hispanic | 1,563,872 | 60,992 | 3.90% |
| American Indian/Alaskan Native (AIAN) | 1,563,872 | 86 | 0.01% |
| Other/Unidentifiable | 1,563,872 | 15,253 | 0.98% |
| *Demographics of Mentees (N, %): Gender* | | | |
| Male | 1,563,872 | 868,546 | 42.32% |
| Female | 1,563,872 | 661,851 | 55.54% |
| Other/Unidentifiable | 1,563,872 | 33,475 | 2.14% |

Table 2. Regressions on Academic Career Success by Intellectual Birth Order

| | **DV: Transition to Advisorship (Cox with Strata)** | | **DV: Students Count (reghdfe)** | | **DV: Log(Short-Term Pubs) (reghdfe)** | | **DV: Log(Long-Term Pubs) (reghdfe)** | |
|---|---|---|---|---|---|---|---|---|
| | **Model 1** | **Model 2** | **Model 3** | **Model 4** | **Model 5** | **Model 6** | **Model 7** | **Model 8** |
| ***Intellectual Birth Order (Ref: First Student)*** | | | | | | | | |
| 2-5 | -0.008 | -0.076$^{**}$ | 0.009 | -0.004 | -0.007$^{**}$ | -0.006+ | -0.006$^{*}$ | -0.003 |
| | (0.015) | (0.026) | (0.010) | (0.016) | (0.002) | (0.003) | (0.002) | (0.004) |
| 6-10 | -0.025 | -0.132$^{***}$ | -0.004 | -0.037+ | -0.017$^{***}$ | -0.009+ | -0.017$^{***}$ | -0.007 |
| | (0.020) | (0.037) | (0.013) | (0.023) | (0.003) | (0.005) | (0.003) | (0.005) |
| 11-15 | -0.060$^{*}$ | -0.182$^{***}$ | -0.026 | -0.082$^{**}$ | -0.022$^{***}$ | -0.011+ | -0.022$^{***}$ | -0.008 |
| | (0.028) | (0.046) | (0.016) | (0.028) | (0.004) | (0.006) | (0.004) | (0.007) |
| 16-20 | -0.059+ | -0.186$^{***}$ | -0.049$^{**}$ | -0.127$^{***}$ | -0.036$^{***}$ | -0.021$^{**}$ | -0.039$^{***}$ | -0.020$^{*}$ |
| | (0.033) | (0.054) | (0.018) | (0.032) | (0.005) | (0.007) | (0.005) | (0.008) |
| 21-30 | -0.119$^{**}$ | -0.226$^{***}$ | -0.032 | -0.128$^{***}$ | -0.044$^{***}$ | -0.021$^{*}$ | -0.048$^{***}$ | -0.021$^{*}$ |
| | (0.041) | (0.064) | (0.021) | (0.037) | (0.005) | (0.008) | (0.006) | (0.009) |
| ≥31 | -0.116$^{*}$ | -0.198$^{*}$ | -0.030 | -0.165$^{***}$ | -0.063$^{***}$ | -0.024$^{*}$ | -0.067$^{***}$ | -0.023$^{*}$ |
| | (0.059) | (0.083) | (0.024) | (0.044) | (0.007) | (0.010) | (0.008) | (0.011) |
| ***Author Race (Ref: White)*** | | | | | | | | |
| Asian/Pacific Islander (API) | 1.033$^{***}$ | 0.888$^{***}$ | 1.116$^{***}$ | 0.994$^{***}$ | 0.076$^{***}$ | 0.045$^{***}$ | 0.109$^{***}$ | 0.075$^{***}$ |
| | (0.012) | (0.015) | (0.011) | (0.013) | (0.002) | (0.002) | (0.002) | (0.003) |
| Black | -0.494$^{***}$ | -0.270$^{**}$ | -0.205$^{***}$ | -0.185$^{***}$ | -0.066$^{***}$ | -0.044$^{***}$ | -0.075$^{***}$ | -0.048$^{***}$ |
| | (0.073) | (0.087) | (0.007) | (0.015) | (0.006) | (0.007) | (0.007) | (0.008) |
| Hispanic | 0.056+ | 0.013 | -0.133$^{***}$ | -0.123$^{***}$ | -0.007+ | -0.010$^{*}$ | 0.002 | -0.002 |
| | (0.028) | (0.035) | (0.006) | (0.010) | (0.004) | (0.004) | (0.004) | (0.005) |
| AIAN | 0.674 | 0.452 | -0.133$^{***}$ | -0.029 | 0.070 | 0.174 | 0.070 | 0.149 |
| | (0.562) | (0.648) | (0.035) | (0.072) | (0.123) | (0.148) | (0.135) | (0.160) |
| Other/Unidentifiable | -0.006 | -0.029 | -0.034$^{**}$ | -0.034 | -0.011+ | -0.022$^{**}$ | -0.004 | -0.015 |
| | (0.064) | (0.075) | (0.012) | (0.018) | (0.007) | (0.008) | (0.007) | (0.009) |
| ***Author Gender (Ref: Male)*** | | | | | | | | |
| Female | -0.168$^{***}$ | -0.113$^{***}$ | -0.192$^{***}$ | -0.175$^{***}$ | -0.073$^{***}$ | -0.045$^{***}$ | -0.077$^{***}$ | -0.049$^{***}$ |
| | (0.011) | (0.014) | (0.006) | (0.008) | (0.001) | (0.002) | (0.002) | (0.002) |
| Other/Unidentifiable | -0.772$^{***}$ | -0.694$^{***}$ | -0.885$^{***}$ | -0.855$^{***}$ | -0.185$^{***}$ | -0.166$^{***}$ | -0.215$^{***}$ | -0.197$^{***}$ |
| | (0.041) | (0.049) | (0.013) | (0.017) | (0.003) | (0.004) | (0.004) | (0.005) |
| ***Mentor/Mentee Abilities*** | | | | | | | | |
| Log(Mentor Experience) | -0.016$^{*}$ | -0.043$^{*}$ | 0.001 | -0.012 | 0.001 | -0.003 | 0.001 | -0.003 |
| | (0.007) | (0.021) | (0.004) | (0.012) | (0.001) | (0.003) | (0.001) | (0.003) |
| Log(Previous Publications) | 0.712$^{***}$ | 0.648$^{***}$ | 0.696$^{***}$ | 0.680$^{***}$ | 1.788$^{***}$ | 1.765$^{***}$ | 1.996$^{***}$ | 1.975$^{***}$ |
| | (0.023) | (0.033) | (0.046) | (0.051) | (0.009) | (0.010) | (0.011) | (0.012) |
| Constant | | | 0.356$^{***}$ | 0.438$^{***}$ | 0.280$^{***}$ | 0.278$^{***}$ | 0.294$^{***}$ | 0.292$^{***}$ |
| | | | (0.008) | (0.020) | (0.002) | (0.004) | (0.002) | (0.005) |
| Intellectual Family Fixed Effects | | YES | | YES | | YES | | YES |
| Year Fixed Effects | YES | YES | YES | YES | YES | YES | YES | YES |
| Field Fixed Effects | YES | YES | YES | YES | YES | YES | YES | YES |
| Adjusted $R^2$ | | | 0.019 | 0.185 | 0.135 | 0.166 | 0.128 | 0.152 |
| $N$ | 1285986 | 1285986 | 1563872 | 1399866 | 1061378 | 940119 | 1061378 | 940119 |

*Note:* $^{*}$ $p < 0.05$, $^{**}$ $p < 0.01$, $^{***}$ $p < 0.001$. *All tests are two-tailed.*

While these numbers may not appear large on their own, the effects of intellectual birth order are comparable in magnitude to widely discussed demographic variables such as gender and race. For example, in the transition to advisorship, the disadvantage of being a female scholar ($\beta$= -0.113$^{***}$) is approximately equivalent to that of students ranked 6–10 within a scholarly family when compared to the firstborn ($\beta$= -0.132$^{***}$; Model 2). Similarly, the reduced likelihood of attracting next generation students as a Hispanic scholar ($\beta$= -0.123$^{***}$) is close to the disadvantage experienced by students ranked 16–20 ($\beta$= -0.127$^{***}$; Model 4).

Figure 3 provides a descriptive illustration of between-sibship and within-sibship differences in outcome variables: another case where inter- and intra-unit variations move in opposite directions. Here, we report average values directly, without additional controls, to uncover the effects absorbed by family fixed effects in the previous regression models. As commonly expected, larger size of scholarly families can serve as a proxy for greater mentor capability (Wang and Shibayama 2022; Wang et al. 2021; Xing et al. 2025). Accordingly, when intellectual birth order is held constant, students from larger scholarly families tend to achieve better career outcomes than those from smaller ones. In other words, a fifth-born student from a large lineage typically performs better than a fifth-born student from a small lineage. Within the same lineage, however, later-born students show a clear tendency to perform worse than their earlier-born siblings on average.

Here, the inter-unit pattern within academic families contrasts with biological families, where wealthier households often have fewer children (Becker and Lewis 1973; Blake 1981; Dribe et al. 2017). Nevertheless, within both biological and intellectual families, the career outcomes of students and the social outcomes of children decline monotonically with birth order, demonstrating similar within-unit patterns.

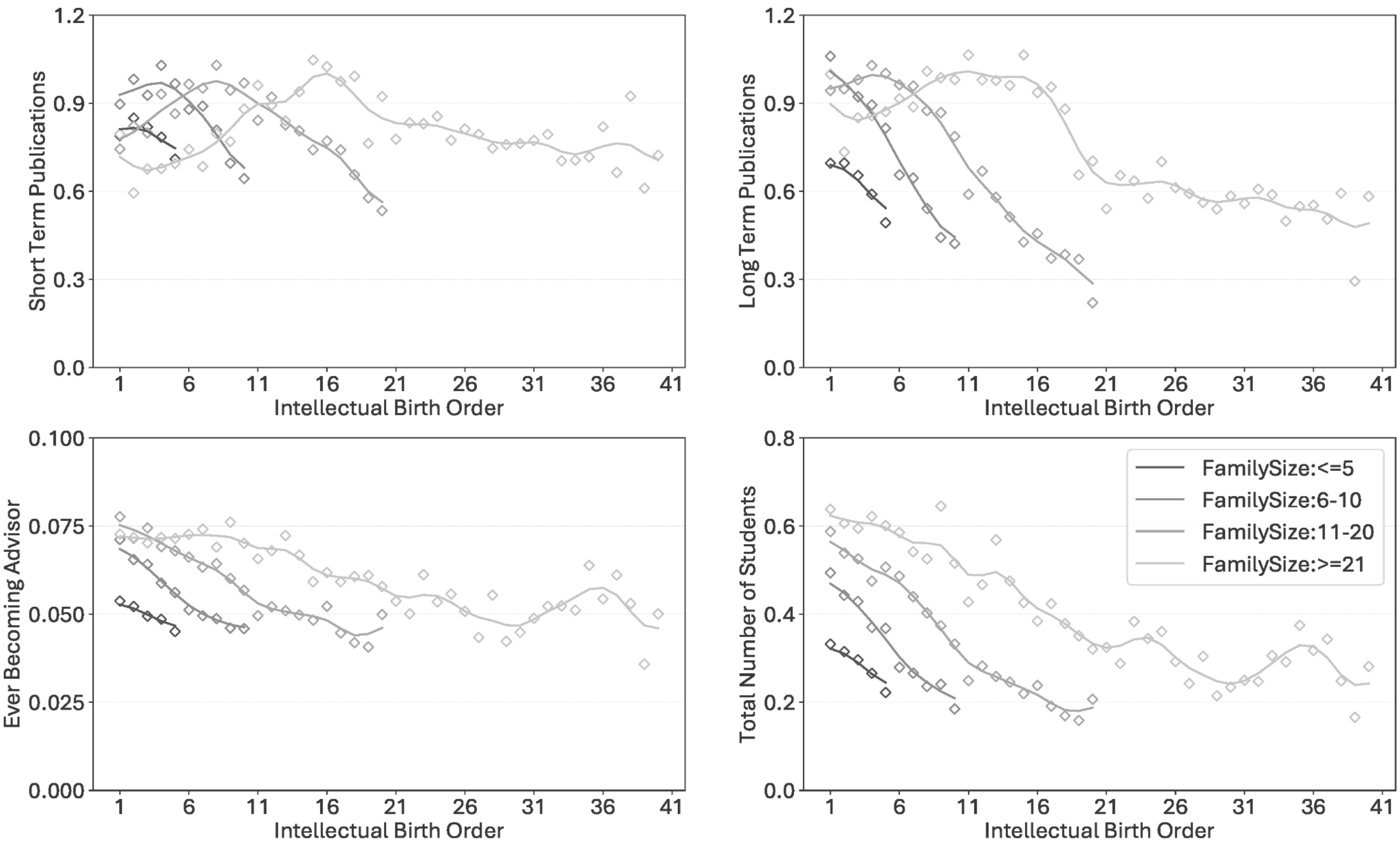

Figure 3. Between- and Within-Sibship Difference of Dependent Variables: A Descriptive Result

***Robustness Test for Causal Interpretation***

The comparability between intellectual and biological families may be limited by a key difference: offspring in intellectual families are often not randomly assigned, but intentionally chosen. Mentors and mentees typically engage in a two-way selection process to determine whether they wish to collaborate in the following years. Consequently, observed birth-order effects may be subject to endogeneity problems and cannot be interpreted as causal. To alleviate this concern, we conduct two further robustness tests: the first one based on the propensity score weighting, and the second based on the permutation test. Both tests are reported in Appendix G and are consistent with the main findings.

### *Competing Explanations*

Our empirical results reveal a clear negative gradient across intellectual birth order. What, then, generates this pattern? The most obvious candidate explanations, including period effects and differences in advisor capabilities or institutional resources, have been fully absorbed by our year and family fixed effects. The mechanism must therefore lie elsewhere. In this section, we examine four competing explanations proposed in existing literature. We define the mechanism variables as outlined in the Data and Method section, and present model results for mechanism analyses in Figure 4, which now reports eleven mechanism panels (a)–(k) covering all four candidate pathways. Details of these models are reported in Appendix B.

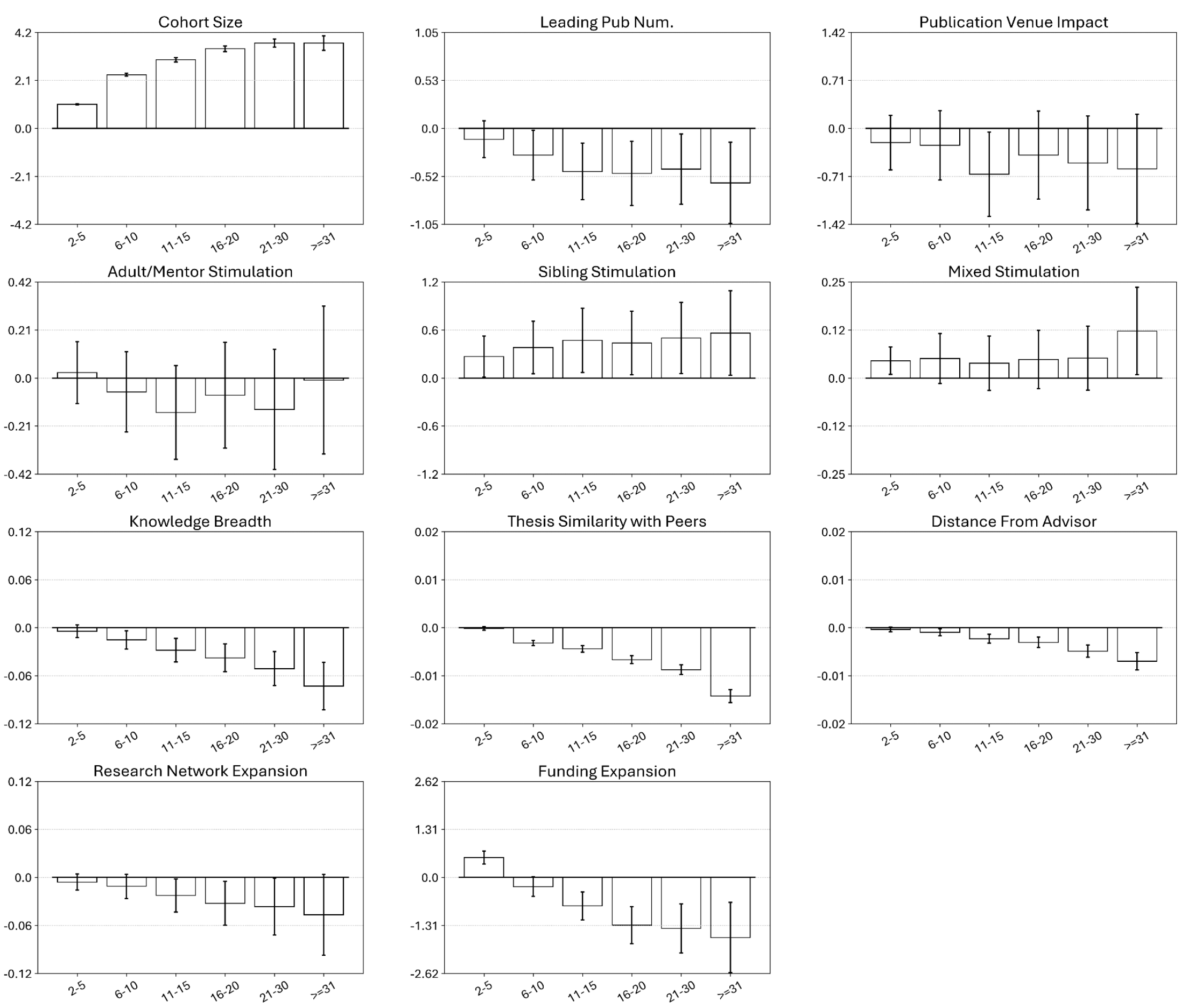

Figure 4. Mechanism Analysis Results

***Resource Dilution Theory.*** Resource dilution theory posits that as the number of children in a family increases, competition for parental resources intensifies, resulting in a smaller share for each child. We test this explanation and report results in the first row of Figure 4. Panels (b) and (c) are based on a subsample of students who produced at least one publication during their Ph.D. studies, ensuring the measures are meaningful.

As shown in panel (a), when we define a cohort as including two years above and two years below the focal student, higher birth order is indeed associated with a larger number of peers. This provides support for the first part of resource dilution theory: later students are more likely to experience a situation in which their mentor supervises a larger number of protégés at the same time, creating a potentially more competitive environment.

Does increasing density within intellectual families lead to a dilution of available resources? In panels (b) and (c) of Figure 4, we use coauthored publications with advisors as a proxy for mentor's investment and attention. While we observe a slight decreasing gradient in opportunities to take leading roles (i.e., first or corresponding authorship), we find no comparable difference in publication impact. Thus, although our empirical results support the descriptive premise of resource dilution theory—that later students enter larger cohorts—they offer no conclusive evidence that these students receive substantially fewer resources than their earlier-born siblings within the same lineage.

While future studies may identify forms of resource dilution beyond publication and academic resources, our results nonetheless echo existing research on the cumulative advantages in scientists' careers. Top-performing scientists early in their careers are more likely to remain high performers throughout their professional trajectories (Merton 1979; Kwiek and Szymula 2025; Zhang et al. 2022; Cole 1973), and their reputation and status can attract exceptional levels of funding, visibility, and access to publication venues for their collaborators and protégés (Azoulay et al. 2010, 2019). A more realistic interpretation, therefore, is that as teams expand, principal investigators also experience a growth in scientific resources as they progress in their careers and accumulate greater capacities, leading to a larger resource pool.

We replicate these models using several alternative measures for robustness tests (See Appendix C), and results remain consistent with those reported here. H2 thus receives only partial support.

***The Confluence Model.*** We next turn to the confluence model. The basic assumption of the confluence model argues that later-borns grow up in a setting of less adult participation and interact more with their siblings (Zajonc and Markus 1975; Zajonc and Mullally 1997).

Results presented in panels (d)–(f) of Figure 4 offer some support for this hypothesis. Among all papers published during the Ph.D. period, the proportion of collaborations with advisors does not systematically differ by birth order: later students appear to have opportunities to collaborate with their advisors comparable to those of earlier students (most coefficients in panel (d) are not significant). Nevertheless, we do observe that later students systematically collaborate more frequently with, and only with, their peers (panel (e)), and this pattern persists even after controlling for cohort size (Table A2, Model 4). Mixed stimulation (Table A2, Models 5 and 6) also show an increasing gradient of coefficients, though none are statistically significant. These findings are consistent with the underlying assumptions of the confluence model, suggesting that later students operate in an intellectual environment characterized by greater peer interaction and relatively less cognitive stimulation from mature scholars.

In Tables A5_1 and A5_2 of Appendix B, we examine pathways linking different types of stimulation to four measures of academic success. Results show that compared with cognitive stimulation from mentors, sibling stimulation exhibits only a faint positive effect on most outcomes. These findings align with the confluence model, suggesting that later students enter an intellectual environment not always conducive to the development of their cognitive abilities. Although we do not find a clear negative causal path between more sibling collaboration and weaker academic achievement, our results indicate that later students face various suppressing factors in their environment that may divert attention from most constructive efforts. Therefore, our empirical results offer partial support for H3.

***Role Specialization*.** Next, we discuss the role-based mechanism: siblings differentiate themselves to reduce competition for parental attention and resources (Hotz and Pantano 2015). We measure the level of knowledge specialization using text data from students' dissertations, including research topics and abstracts. We rely on dissertations rather than publications because dissertations are primarily individual-led projects, whereas publications are more likely influenced by collaboration. As such, dissertations provide a more accurate reflection of a Ph.D. scholar's intellectual interests and identities.

Figure 4 panels (g)–(i) provide support for the hypothesis of role specialization. The results from panels (g) and (h) show a very clear pattern, that later students consistently exhibit a narrower scope of exploration in their dissertations (significant, negative coefficients with a decreasing gradient), and that later students do display greater differentiation from peers. In panel (i), we further extend this analysis to differentiation from the advisor. Interestingly, later students' dissertations are systematically closer to their advisor's research, with the coefficient on cosine distance growing increasingly negative with birth order. This pattern reveals a distinctive feature of role specialization within academic lineages: role differentiation operates horizontally against peers, but not vertically against the advisor. Earlier students enter an expanding lineage and can claim positions at its periphery or beyond. Later students inherit a sub-line of an established research program and must operate within its boundaries, carving out narrow positions relative to a growing cohort of peers (Panel a).

Our results may seem to contradict prior studies that emphasize the value of specialization in doctoral training (Heiberger et al. 2021). However, the tension dissolves when we consider the level of comparison. Prior work compares specialization across individuals and indexes knowledge depth and returns to expertise. Here we compare within lineage, where specialization reflects both the intellectual taste and the position a student is allocated within an internal division of labor. Appendix Tables A5_1 and A5_2 confirm this reading. Over-specialization from peers is associated with poorer outcomes, and distance from the advisor follows a U-shape: very close (protected successors) and very far (independent scholars) both outperform intermediate positions. Later students are systematically pushed toward the lower ends of the curve, close enough to compete with many peers for the protected niche, and not far enough to claim independence.

Thus, the effects identified in our models consistently indicate the presence of role specialization for later students in academic families, and support H4, suggesting that as academic lineages expand, students increasingly differentiate their research topics to establish distinct intellectual identities, but these dynamics may limit them from building independent identities.

***Family Dynamics Perspective.*** Beyond differentiation among peers, the family dynamics perspective predicts that the lab environment might have shifted when later students join. We test this hypothesis using two lab-level proxies measured at the focal student's entry, reported in panels (j)–(k) of Figure 4. The pattern is consistent across both variables. In panel (j), Research Network Expansion shows a small but systematic decline with intellectual birth order: later students enter lineages whose coauthorship networks are no longer growing at the rate they were when earlier students arrived. Panel (k) reveals an inverted-U pattern in Funding Expansion: the lab's funding base expands first during the early stages but later plateaus and contracts as the lineage matures. By the time later-born students join, the lineage’s funding base has entered a more settled phase. Taken together, these findings support H5, and suggest that lineage age and lineage stage operate as a distinct mechanism. The lab a tenth student joins can differ substantially from the one firstborns encountered, carrying different priorities and placing different demands on its members.

We summarize the pathways in Figure 5 below.

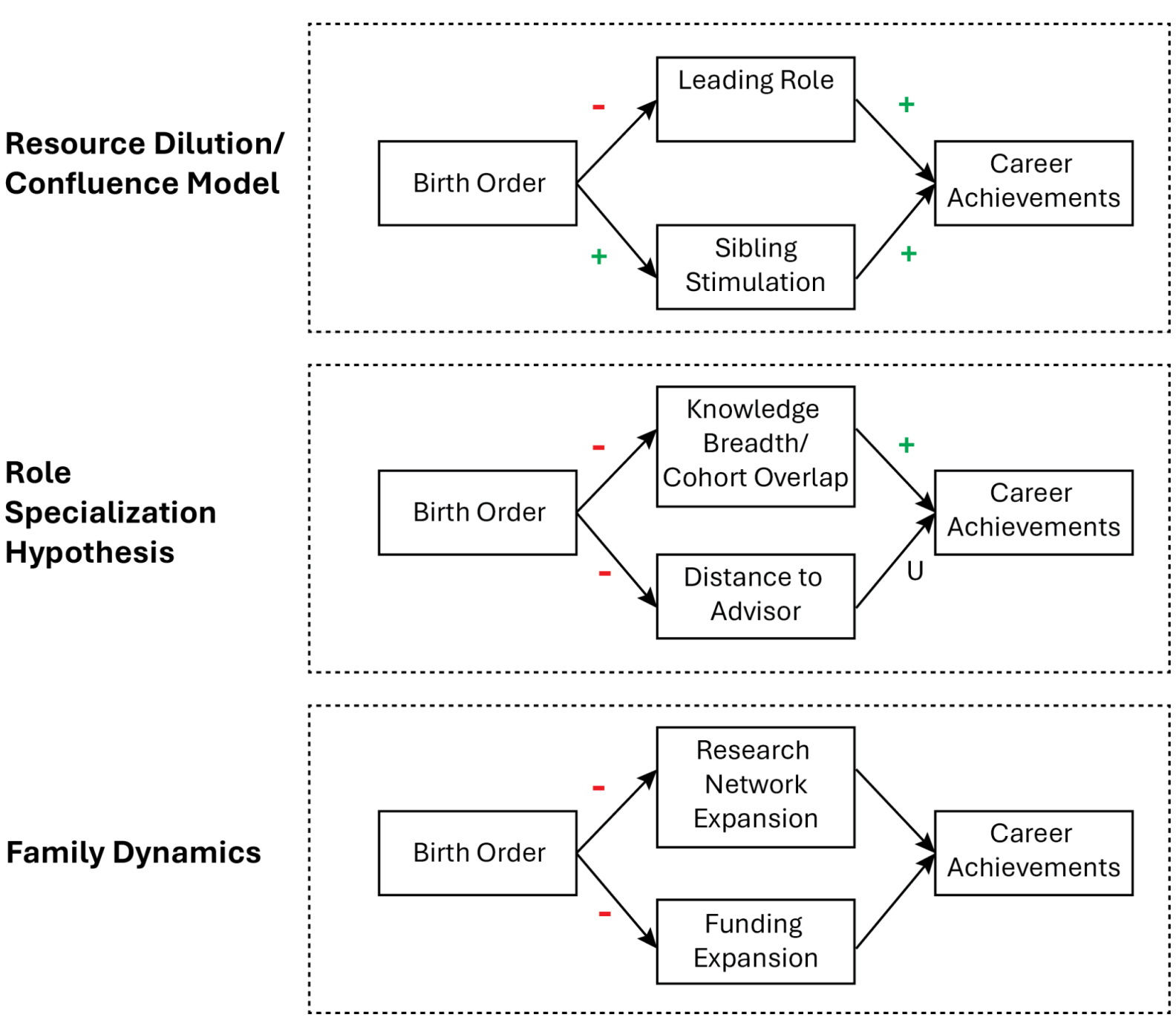

Figure 5. Mechanisms for Intellectual Birth Order Effects

### *Cross-Sample Heterogeneity*

Finally, we assess how the effects of intellectual birth order differ across subsamples of the dataset: students of different gender (male vs. female), students of different race (white vs. other races), university type (R1 vs. non-R1), and discipline (STEM vs. others). We re-estimate the same models within each subsample and report the conditional marginal effects in Figure 6. Detailed model results are provided in Appendix B (Tables A6_1–A6_4).

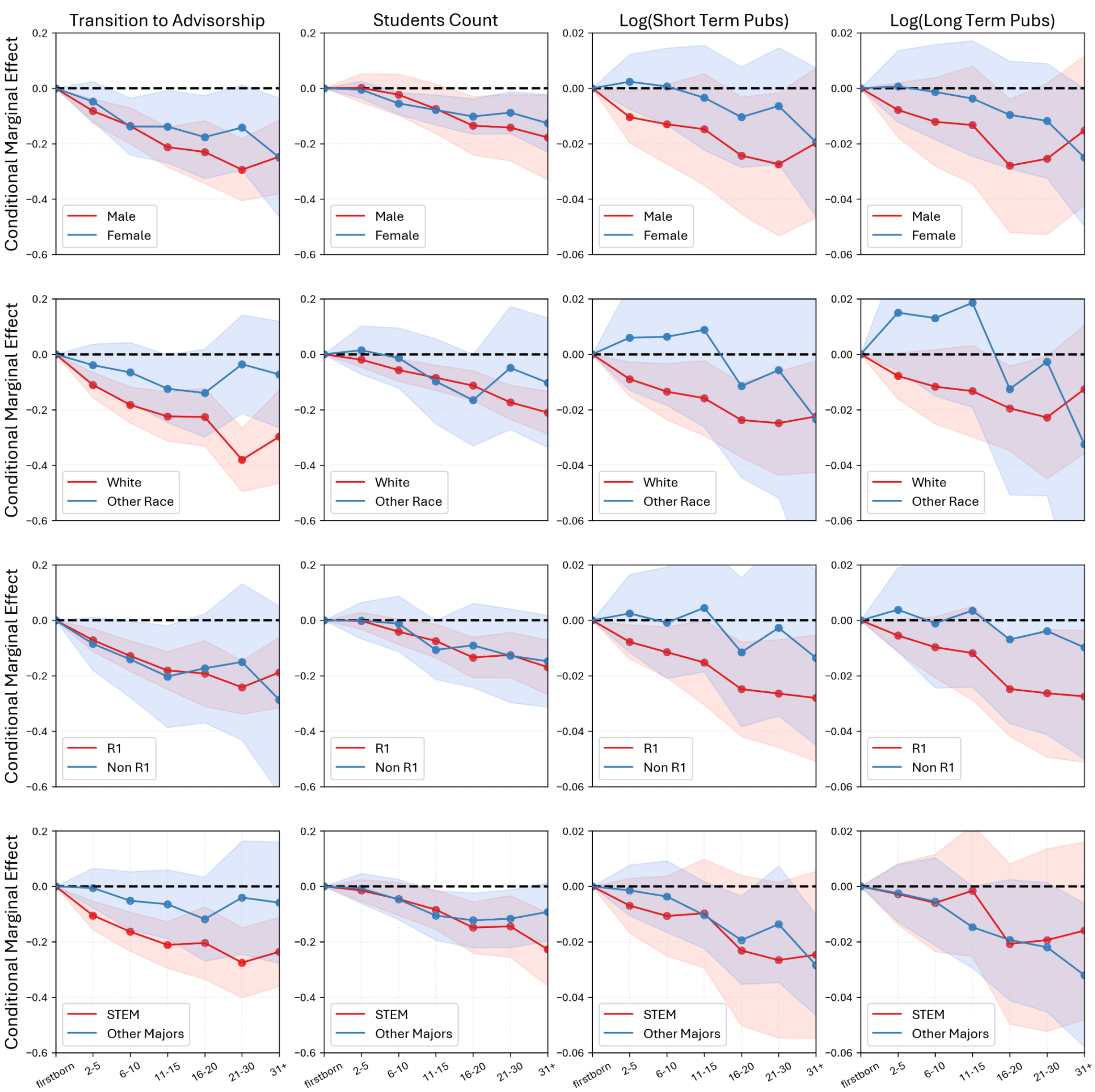

Figure 6. Cross-Sample Heterogeneity in the Effects of Intellectual Birth Order

Our results reveal systematic cross-group heterogeneity. As shown in Figure 6, the effects of intellectual birth order, and the firstborn premium they imply, are most pronounced among students in R1 universities and STEM fields, among male students, and among white students. Given the systematic disparities between groups, the muted birth order effects for female, minority, and non-R1 students could be read as evidence of structural disadvantages that prevent them from realizing the firstborn premium even when they occupy firstborn positions.

## Discussion and Conclusion

This analysis integrates big data analytics with a demographic perspective to revisit a classic question in the social study of science: how the scientific profession is stratified (Merton 1979; Keith and Babchuk 1998; Heiberger et al. 2021). While acknowledging the enduring significance of the structural Matthew effect across multiple levels of analysis, we argue that this perspective may be increasingly limited in today's transforming academic landscape, where the individual-based, elite model of science is gradually giving way to new forms of team science and collective intelligence. Consequently, we shift our attention from variation across institutions and intellectual lineages to variation within the same institution and lineage, where structural power weakens and relational and interactive dynamics become more prominent. This new research setting calls for a demographic framework grounded in striking parallel between biological and intellectual reproduction across generations, as we have discussed in detail.

As a first step and illustration of this agenda, this paper adopts the demographic concept of birth order to examine how students within the same institutional and intellectual context may systematically experience different career opportunities and outcomes. Our findings highlight the striking parallel between biological and intellectual systems in their reproductive mechanisms. As in biological families, later students in a scholarly lineage tend to perform worse than earlier ones on average across all four outcome measures. Practically, this does not imply that students should prefer early positions in young labs to later positions in established ones. When inter-lineage variation is accounted for, the tenth student of a renowned scientist may still outperform, on average, the first student of a less established scholar who supervises fewer students over the course of their career (Figure 3). Nevertheless, our results indicate that using an advisor's early students as a benchmark for prediction may also be overly optimistic, as early students tend to systematically achieve better outcomes than those who come later. Whether a student can secure the most advantageous position, the earliest student of a successful scholar, thus becomes a matter of luck and risk.

Our findings should not be read as a critique of established labs or their principal investigators. The within-lineage patterns we document reflect structural features of academic life cycles that operate regardless of advisor quality or lab effectiveness. The patterns we reveal are average estimates from regression models. At the individual level, established labs often produce later students whose absolute outcomes exceed those of their earlier-born peers.

This project cross-fertilizes sociological theories, contributing to the sociology of science, demography, and theories of social stratification. For the social study of science, it moves beyond the traditional structural framework in explaining scientific stratification, offering a new perspective that conceptualizes knowledge production as a population process. While this paper focuses only on intellectual birth order, the proposed theoretical lens can be extended to other key social dynamics within scientific institutions, such as collaboration, professional aging, and career termination, which can all examined through a demographic perspective (Cui et al. 2025).

For demographic research, this study illustrates the potential of applying well-established demographic theories and methodological tools to other social systems. Although family and population processes and professional processes have traditionally been examined in separate subfields of sociology, many of their underlying dynamics are analogous and shaped by similar mechanisms. For example, our findings suggest that cognitive stimulation plays a crucial role both in children's development within biological families and in scholars' growth within intellectual lineages. Extending demographic theories and methods to other social and organizational domains can therefore catalyze new empirical insight.

Like all research, this project has its limitations. First, because our analysis relies on bibliometric data rather than close qualitative observations of mentor–mentee interactions, it may not capture every dimension of birth-order effects. This issue applies to both mechanism and career variables. Thus, if we find no evidence supporting a mechanism, we cannot conclude that the mechanism is absent, but only that it is not observable within our current dataset and resolution of measurement. For the same reason, our findings may more accurately reflect patterns in STEM disciplines, where coauthored publications are common, than in the humanities and interpretive social sciences, where publication practices and mentorship relations are less tightly coupled. We therefore call for future research to collect richer data, especially observational and qualitative evidence, to explore this topic in greater depth. Second, our data do not permit strict causal inference. Although we have incorporated high-dimensional fixed effects for both lineage and year and conducted multiple robustness checks, including propensity score weighting and permutation tests, to ensure that the identified patterns are stable and meaningful, some selection biases may still remain unaccounted for. Future studies could employ more rigorous research designs, such as natural experiments, to disentangle selection effects from treatment effects in the study of intellectual birth-order patterns.

Despite these limitations, we envision fruitful extensions and applications of the framework, both in theory and in practice. Theoretically, our work has rich implications for sociological theory, but it also suggests important applications and policies. The first and most direct implications of our findings concern innovation policy. Our results show that innovation occurs locally. Many dimensions emphasized in prior research, such as specialization, are not merely reflections of intellectual preference or historical trends, but also outcomes of the organizational structure of scientific institutions. Second, our work offers insights for scientific education. By highlighting the systematic differences in the intellectual environments that early and later students encounter, our study underscores the trade-off between financial and intellectual resources students navigate as they undertake their career journey.

The expansion of science has transformed academic stratification from a problem of institutional selection into one of organizational dynamics. Where the structuralist tradition locates cumulative advantage between lineages, we find systematic inequality emerging within them through the everyday rhythms of cohort entry, peer collaboration, and intellectual differentiation. Birth order in scholarly families, like birth order in biological ones, marks not just sequence but life chances. This parallel reveals stratification as simultaneously structural and interactional, shaped by both the distribution of resources across institutions and the micro-politics of labs and seminar rooms. As science continues its shift toward mass production and team-based work, understanding inequality will require attending not only to who enters which institutions, but to how advantage and disadvantage are produced in the daily practice of knowledge work itself.

# Appendices

## Appendix A. Procedure for Matching the ProQuest Dissertations with OpenAlex Publication Records

We merge the proquest dissertation dataset with the OpenAlex data, following the procedures proposed by Heiberger and colleagues (Heiberger et al. 2021). Detailed steps are as follows:

1. **Name Disambiguation for OpenAlex:** OpenAlex has a well-known data-quality issue in which a single researcher is often fragmented across multiple author IDs (predominantly many-to-one, with some many-to-many cases) (Zhou and Sun 2024). We therefore disambiguate authors before matching OpenAlex records against ProQuest. We first train a word2vec model on authors' publication records, and locate each author ID in semantic space as the mean vector of its publications. We then group all records sharing the same name, and run the DBSCAN clustering algorithm on their semantic vectors; same-name records that fall into closely clustered semantic regions are treated as a single researcher and assigned a new, unified ID. Steps 2–4 of the name-matching procedure are keyed on these reassigned IDs.
2. **Name matching:** We begin by performing a fuzzy match between author/advisor names in ProQuest and OpenAlex, retaining multiple candidate matches for each OpenAlex author to account for minor typographical variations.
3. **Temporal filtering:** Using the graduation dates provided in ProQuest, we restrict OpenAlex matches to individuals with a plausible publication timeline, excluding candidates whose nearest publication appears more than ten years before or after their recorded graduation year.
4. **Content validation:** Finally, we compare the textual content of dissertations and publications for all candidate pairs to identify the most accurate match. To do so, we train a word2vec model to compute semantic similarity between (1) each author's dissertation abstract and (2) the abstracts of their candidate publications in OpenAlex. The pair with the highest cosine similarity score is selected as the final match. We use the average of their students' dissertation vectors to represent advisors' positions in the ProQuest dataset.

## Appendix B. Regression Models for Figures in the Main Text

In this Appendix, we present the detailed models underlying the visualizations in the main text. We begin by outlining the model specifications used to generate the results shown in Figure 4 (mechanism analysis). The corresponding results are reported in Tables A1–A4 below.

In Table A5_1 and Table A5_2, we further test whether increased resources and cognitive stimulation, greater knowledge specialization and changing family dynamics are associated with poorer academic performance. The results are significant, and the coefficient directions align with expectations, supporting the theoretical arguments of the confluence model and the role specialization hypothesis. Table A5 serves as the foundation of Figure 5.

Finally, in Tables A6_1–A6_4, we examine whether the effects of intellectual birth order vary across theoretically meaningful subgroups: students' gender (male vs. female), race (white vs. nonwhite), university type (R1 vs. non-R1), and disciplinary domain (STEM vs. non-STEM). For each dimension, we re-estimate the models separately within each subgroup and report the corresponding marginal effects.

Table A1. Mechanism 1: Resource Dilution Theory

| | DV: Cohort Size | DV: Number of Leading Papers with Advisor | | DV: Average Impact Factor of Publication Channel | |
|---|---|---|---|---|---|
| | **Model 1** | **Model 2** | **Model 3** | **Model 4** | **Model 5** |
| ***Intellectual Birth Order (Ref: First Student)*** | | | | | |
| 2-5 | 1.052$^{***}$ | -0.119 | -0.102 | -0.212 | -0.210 |
| | (0.015) | (0.102) | (0.104) | (0.207) | (0.210) |
| 6-10 | 2.343$^{***}$ | -0.292$^{*}$ | -0.257+ | -0.251 | -0.249 |
| | (0.030) | (0.139) | (0.144) | (0.262) | (0.268) |
| 11-15 | 3.002$^{***}$ | -0.471$^{**}$ | -0.425$^{**}$ | -0.681$^{*}$ | -0.677$^{*}$ |
| | (0.048) | (0.157) | (0.163) | (0.319) | (0.319) |
| 16-20 | 3.480$^{***}$ | -0.493$^{**}$ | -0.441$^{*}$ | -0.395 | -0.391 |
| | (0.064) | (0.180) | (0.186) | (0.333) | (0.349) |
| 21-30 | 3.732$^{***}$ | -0.445$^{*}$ | -0.389+ | -0.512 | -0.508 |
| | (0.091) | (0.196) | (0.201) | (0.356) | (0.363) |
| >=31 | 3.729$^{***}$ | -0.596$^{**}$ | -0.534$^{*}$ | -0.603 | -0.598 |
| | (0.160) | (0.227) | (0.232) | (0.413) | (0.422) |
| Cohort Size | | | -0.013 | | -0.001 |
| | | | (0.009) | | (0.015) |
| ***Author Race (Ref: White)*** | | | | | |
| Asian/Pacific Islander (API) | -0.002 | 0.117$^{*}$ | 0.117$^{*}$ | -0.131 | -0.131 |
| | (0.006) | (0.051) | (0.051) | (0.083) | (0.083) |
| Black | 0.003 | -0.154 | -0.154 | -0.205 | -0.205 |
| | (0.023) | (0.157) | (0.157) | (0.376) | (0.376) |
| Hispanic | 0.037$^{**}$ | -0.454$^{***}$ | -0.455$^{***}$ | 0.171 | 0.171 |
| | (0.011) | (0.080) | (0.080) | (0.209) | (0.209) |
| AIAN | -0.186 | 0.598 | 0.614 | -1.311 | -1.310 |
| | (0.286) | (0.845) | (0.844) | (1.328) | (1.326) |
| Other/Unidentifiable | 0.013 | 0.131 | 0.131 | 0.244 | 0.244 |
| | (0.020) | (0.154) | (0.154) | (0.337) | (0.337) |
| ***Author Gender (Ref: Male)*** | | | | | |
| Female | 0.001 | -0.195$^{***}$ | -0.195$^{***}$ | -0.247$^{***}$ | -0.247$^{***}$ |
| | (0.004) | (0.042) | (0.042) | (0.073) | (0.073) |
| Other/Unidentifiable | -0.028+ | -0.305$^{*}$ | -0.306$^{*}$ | 0.166 | 0.166 |
| | (0.017) | (0.142) | (0.142) | (0.294) | (0.294) |
| Log(Mentor Experience) | 0.563$^{***}$ | -0.174$^{*}$ | -0.165$^{*}$ | -0.039 | -0.038 |
| | (0.017) | (0.069) | (0.070) | (0.132) | (0.132) |
| Log(Previous Publications) | 0.026$^{*}$ | 8.857$^{***}$ | 8.858$^{***}$ | 0.587$^{***}$ | 0.587$^{***}$ |
| | (0.012) | (0.365) | (0.365) | (0.079) | (0.079) |
| Constant | 2.580$^{***}$ | 2.920$^{***}$ | 2.949$^{***}$ | 3.885$^{***}$ | 3.887$^{***}$ |
| | (0.043) | (0.150) | (0.149) | (0.228) | (0.231) |
| Intellectual Family Fixed Effects | YES | YES | YES | YES | YES |
| Year Fixed Effects | YES | YES | YES | YES | YES |
| Field Fixed Effects | YES | YES | YES | YES | YES |
| Adjusted $R^2$ | 0.820 | 0.227 | 0.227 | 0.060 | 0.060 |
| *N* | 1399866 | 215249 | 215249 | 139817 | 139817 |

*Note:* $^{*}$ $p < 0.05$, $^{**}$ $p < 0.01$, $^{***}$ $p < 0.001$. *All tests are two-tailed.*

Table A2. Mechanism 2: The Confluence Model

| | **DV: Adult Stimulation (Num. of Papers with Advisor but not Siblings)** | | **DV: Sibling Stimulation (Num. of Papers with Siblings but not Advisor)** | | **DV: Mixed Stimulation (Num. of Papers with both Advisor and Siblings)** | |
|---|---|---|---|---|---|---|
| | **Model 1** | **Model 2** | **Model 3** | **Model 4** | **Model 5** | **Model 6** |
| ***Intellectual Birth Order (Ref: First Student)*** | | | | | | |
| 2-5 | 0.024 | 0.039 | 0.268$^{*}$ | 0.254+ | 0.045$^{*}$ | 0.030 |
| | (0.070) | (0.071) | (0.130) | (0.132) | (0.018) | (0.019) |
| 6-10 | -0.062 | -0.030 | 0.381$^{*}$ | 0.352$^{*}$ | 0.051 | 0.019 |
| | (0.091) | (0.094) | (0.168) | (0.173) | (0.033) | (0.036) |
| 11-15 | -0.153 | -0.111 | 0.470$^{*}$ | 0.433$^{*}$ | 0.038 | -0.002 |
| | (0.106) | (0.110) | (0.205) | (0.206) | (0.036) | (0.036) |
| 16-20 | -0.076 | -0.029 | 0.437$^{*}$ | 0.394+ | 0.048 | 0.002 |
| | (0.120) | (0.125) | (0.203) | (0.209) | (0.039) | (0.041) |
| 21-30 | -0.139 | -0.088 | 0.501$^{*}$ | 0.455+ | 0.052 | 0.002 |
| | (0.136) | (0.141) | (0.227) | (0.232) | (0.043) | (0.044) |
| ≥31 | -0.009 | 0.046 | 0.563$^{*}$ | 0.513+ | 0.123$^{*}$ | 0.069 |
| | (0.167) | (0.171) | (0.270) | (0.273) | (0.058) | (0.060) |
| Cohort Size | | -0.012$^{*}$ | | 0.011$^{*}$ | | 0.011$^{***}$ |
| | | (0.005) | | (0.005) | | (0.002) |
| ***Author Race (Ref: White)*** | | | | | | |
| Asian/Pacific Islander (API) | -0.495$^{***}$ | -0.495$^{***}$ | 0.072 | 0.072 | -0.050$^{*}$ | -0.050$^{*}$ |
| | (0.035) | (0.035) | (0.096) | (0.096) | (0.020) | (0.021) |
| Black | -0.152$^{*}$ | -0.153$^{*}$ | 0.010 | 0.011 | -0.046+ | -0.045+ |
| | (0.073) | (0.073) | (0.059) | (0.059) | (0.024) | (0.024) |
| Hispanic | -0.396$^{***}$ | -0.396$^{***}$ | 0.036 | 0.036 | -0.005 | -0.005 |
| | (0.072) | (0.072) | (0.056) | (0.056) | (0.026) | (0.026) |
| AIAN | 0.338 | 0.344 | -0.142+ | -0.147+ | -0.021 | -0.026 |
| | (0.404) | (0.406) | (0.076) | (0.076) | (0.029) | (0.030) |
| Other/Unidentifiable | 0.204 | 0.205 | 0.002 | 0.002 | -0.036 | -0.037 |
| | (0.126) | (0.126) | (0.064) | (0.064) | (0.023) | (0.023) |
| ***Author Gender (Ref: Male)*** | | | | | | |
| Female | -0.038 | -0.037 | 0.080 | 0.080 | -0.021$^{*}$ | -0.021$^{*}$ |
| | (0.028) | (0.028) | (0.069) | (0.069) | (0.010) | (0.010) |
| Other/Unidentifiable | 0.647$^{**}$ | 0.647$^{**}$ | -0.427 | -0.427 | 0.038 | 0.038 |
| | (0.205) | (0.205) | (0.468) | (0.468) | (0.047) | (0.047) |
| Log(Mentor Experience) | -0.146$^{**}$ | -0.139$^{**}$ | -0.039 | -0.046 | -0.008 | -0.015 |
| | (0.053) | (0.053) | (0.061) | (0.060) | (0.014) | (0.014) |
| Log(Previous Publications) | 0.759$^{***}$ | 0.759$^{***}$ | 0.494$^{*}$ | 0.493$^{*}$ | 0.172$^{*}$ | 0.172$^{*}$ |
| | (0.142) | (0.142) | (0.225) | (0.225) | (0.083) | (0.083) |
| Constant | 1.720$^{***}$ | 1.747$^{***}$ | -0.141 | -0.165 | 0.138$^{***}$ | 0.112$^{***}$ |
| | (0.102) | (0.101) | (0.190) | (0.188) | (0.026) | (0.027) |
| Intellectual Family Fixed Effects | YES | YES | YES | YES | YES | YES |
| Year Fixed Effects | YES | YES | YES | YES | YES | YES |
| Field Fixed Effects | YES | YES | YES | YES | YES | YES |
| Adjusted $R^2$ | 0.200 | 0.200 | 0.462 | 0.462 | 0.370 | 0.370 |
| *N* | 215249 | 215249 | 215249 | 215249 | 215249 | 215249 |

*Note:* $^{*}$ *p < 0.05,* $^{**}$ *p < 0.01,* $^{***}$ *p < 0.001. All tests are two-tailed.*

Table A3. Mechanism 3: The Explanation based on Role Specialization

| | **DV: Knowledge Breadth in Dissertation (Subject Level)** | | **DV: Dissertation Similarity with Cohorts (Token Level Jaccard)** | | **DV: Distance from Advisor (Cosine Distance between Thesis and Advisor Pub)** | |
|---|---|---|---|---|---|---|
| | **Model 1** | **Model 2** | **Model 3** | **Model 4** | **Model 5** | **Model 6** |
| ***Intellectual Birth Order (Ref: First Student)*** | | | | | | |
| 2-5 | -0.005 | -0.004 | -0.000 | 0.001$^{***}$ | -0.000 | -0.000 |
| | (0.004) | (0.004) | (0.000) | (0.000) | (0.000) | (0.000) |
| 6-10 | -0.016$^{**}$ | -0.014$^{*}$ | -0.004$^{***}$ | -0.000 | -0.001$^{**}$ | -0.001$^{*}$ |
| | (0.006) | (0.006) | (0.000) | (0.000) | (0.000) | (0.000) |
| 11-15 | -0.029$^{***}$ | -0.027$^{***}$ | -0.006$^{***}$ | -0.001 | -0.003$^{***}$ | -0.003$^{***}$ |
| | (0.008) | (0.008) | (0.000) | (0.000) | (0.001) | (0.001) |
| 16-20 | -0.039$^{***}$ | -0.037$^{***}$ | -0.008$^{***}$ | -0.003$^{***}$ | -0.004$^{***}$ | -0.004$^{***}$ |
| | (0.009) | (0.010) | (0.001) | (0.001) | (0.001) | (0.001) |
| 21-30 | -0.053$^{***}$ | -0.051$^{***}$ | -0.011$^{***}$ | -0.005$^{***}$ | -0.006$^{***}$ | -0.006$^{***}$ |
| | (0.011) | (0.011) | (0.001) | (0.001) | (0.001) | (0.001) |
| ≥31 | -0.076$^{***}$ | -0.074$^{***}$ | -0.018$^{***}$ | -0.011$^{***}$ | -0.009$^{***}$ | -0.009$^{***}$ |
| | (0.016) | (0.016) | (0.001) | (0.001) | (0.001) | (0.001) |
| Cohort Size | | -0.001 | | -0.002$^{***}$ | | -0.000 |
| | | (0.001) | | (0.000) | | (0.000) |
| ***Author Race (Ref: White)*** | | | | | | |
| Asian/Pacific Islander (API) | -0.005+ | -0.005+ | 0.001$^{***}$ | 0.001$^{***}$ | -0.009$^{***}$ | -0.009$^{***}$ |
| | (0.003) | (0.003) | (0.000) | (0.000) | (0.000) | (0.000) |
| Black | -0.001 | -0.001 | 0.000 | 0.000 | -0.001 | -0.001 |
| | (0.011) | (0.011) | (0.000) | (0.000) | (0.001) | (0.001) |
| Hispanic | 0.010+ | 0.010+ | 0.000 | 0.000 | -0.001 | -0.001 |
| | (0.005) | (0.005) | (0.000) | (0.000) | (0.000) | (0.000) |
| AIAN | 0.195 | 0.195 | -0.004 | -0.004 | 0.002 | 0.002 |
| | (0.142) | (0.141) | (0.006) | (0.006) | (0.010) | (0.010) |
| Other/Unidentifiable | -0.002 | -0.002 | 0.000 | 0.000 | -0.004$^{***}$ | -0.004$^{***}$ |
| | (0.010) | (0.010) | (0.000) | (0.000) | (0.001) | (0.001) |
| ***Author Gender (Ref: Male)*** | | | | | | |
| Female | -0.008$^{***}$ | -0.008$^{***}$ | 0.000$^{***}$ | 0.000$^{***}$ | -0.001$^{***}$ | -0.001$^{***}$ |
| | (0.002) | (0.002) | (0.000) | (0.000) | (0.000) | (0.000) |
| Other/Unidentifiable | -0.009 | -0.009 | 0.000 | 0.000 | -0.002$^{***}$ | -0.002$^{***}$ |
| | (0.007) | (0.007) | (0.000) | (0.000) | (0.001) | (0.001) |
| Log(Mentor Experience) | -0.001 | -0.001 | -0.005$^{***}$ | -0.003$^{***}$ | -0.000+ | -0.000+ |
| | (0.004) | (0.004) | (0.000) | (0.000) | (0.000) | (0.000) |
| Log(Previous Publications) | -0.005 | -0.005 | 0.000 | 0.000 | -0.002$^{***}$ | -0.002$^{***}$ |
| | (0.006) | (0.006) | (0.000) | (0.000) | (0.000) | (0.000) |
| Constant | 2.070$^{***}$ | 2.072$^{***}$ | 0.160$^{***}$ | 0.165$^{***}$ | 0.408$^{***}$ | 0.408$^{***}$ |
| | (0.006) | (0.006) | (0.000) | (0.000) | (0.000) | (0.000) |
| Intellectual Family Fixed Effects | YES | YES | YES | YES | YES | YES |
| Year Fixed Effects | YES | YES | YES | YES | YES | YES |
| Field Fixed Effects | YES | YES | YES | YES | YES | YES |
| Adjusted $R^2$ | 0.581 | 0.581 | 0.715 | 0.718 | 0.782 | 0.782 |
| $N$ | 1399866 | 1399866 | 1258930 | 1258930 | 1327498 | 1327498 |

*Note:* $^{*}$ $p < 0.05$, $^{**}$ $p < 0.01$, $^{***}$ $p < 0.001$. *All tests are two-tailed.*

Table A4. Mechanism 4: Family Dynamics Model

| | DV: Research Network Expansion Speed | | DV: Funding Expansion Speed | |
|---|---|---|---|---|
| | **Model 1** | **Model 2** | **Model 3** | **Model 4** |
| ***Intellectual Birth Order (Ref: First Student)*** | | | | |
| 2-5 | -0.006 | -0.007 | 0.542$^{***}$ | 0.402$^{***}$ |
| | (0.005) | (0.006) | (0.090) | (0.091) |
| 6-10 | -0.012 | -0.014+ | -0.253 | -0.539$^{***}$ |
| | (0.008) | (0.009) | (0.138) | (0.139) |
| 11-15 | -0.024$^{*}$ | -0.027$^{*}$ | -0.781$^{***}$ | -1.128$^{***}$ |
| | (0.011) | (0.012) | (0.194) | (0.192) |
| 16-20 | -0.034$^{*}$ | -0.037$^{*}$ | -1.308$^{***}$ | -1.688$^{***}$ |
| | (0.015) | (0.015) | (0.259) | (0.254) |
| 21-30 | -0.038$^{*}$ | -0.042$^{*}$ | -1.398$^{***}$ | -1.763$^{***}$ |
| | (0.019) | (0.019) | (0.342) | (0.334) |
| ≥31 | -0.049+ | -0.053$^{*}$ | -1.646$^{***}$ | -1.918$^{***}$ |
| | (0.027) | (0.026) | (0.491) | (0.474) |
| Cohort Size | | 0.001 | | 0.148$^{***}$ |
| | | (0.001) | | (0.022) |
| ***Author Race (Ref: White)*** | | | | |
| Asian/Pacific Islander (API) | -0.003 | -0.003 | -0.005 | -0.004 |
| | (0.002) | (0.002) | (0.036) | (0.036) |
| Black | -0.002 | -0.002 | -0.232 | -0.230 |
| | (0.010) | (0.010) | (0.144) | (0.144) |
| Hispanic | 0.006 | 0.006 | 0.023 | 0.020 |
| | (0.005) | (0.005) | (0.075) | (0.075) |
| AIAN | -0.168 | -0.168 | -2.456 | -2.416 |
| | (0.122) | (0.122) | (1.951) | (1.921) |
| Other/Unidentifiable | 0.001 | 0.001 | 0.218 | 0.217 |
| | (0.009) | (0.009) | (0.147) | (0.147) |
| ***Author Gender (Ref: Male)*** | | | | |
| Female | -0.002 | -0.002 | -0.008 | -0.009 |
| | (0.002) | (0.002) | (0.031) | (0.031) |
| Other/Unidentifiable | -0.009 | -0.009 | 0.152 | 0.155 |
| | (0.007) | (0.007) | (0.124) | (0.124) |
| Log(Mentor Experience) | -0.015$^{**}$ | -0.015$^{**}$ | -1.841$^{***}$ | -1.939$^{***}$ |
| | (0.005) | (0.005) | (0.093) | (0.093) |
| Log(Previous Publications) | -0.003 | -0.003 | -0.029 | -0.034 |
| | (0.005) | (0.005) | (0.070) | (0.070) |
| Constant | 0.383$^{***}$ | 0.380$^{***}$ | 14.471$^{***}$ | 14.111$^{***}$ |
| | (0.009) | (0.010) | (0.195) | (0.207) |
| Intellectual Family Fixed Effects | YES | YES | YES | YES |
| Year Fixed Effects | YES | YES | YES | YES |
| Field Fixed Effects | YES | YES | YES | YES |
| Adjusted $R^2$ | 0.280 | 0.280 | 0.539 | 0.540 |
| *N* | 1156354 | 1156354 | 286819 | 286819 |

*Note:* $^{*}$ *$p < 0.05$,* $^{**}$ *$p < 0.01$,* $^{***}$ *$p < 0.001$. All tests are two-tailed.*

Table A5_1. Effect Pathways from Intellectual Birth Order to Career Outcomes (Advisorship)

| | DV: Transition to Advisorship (Cox with Strata) | | | DV: Students Count (reghdfe) | | |
|---|---|---|---|---|---|---|
| | **Model 1** | **Model 2** | **Model 3** | **Model 4** | **Model 5** | **Model 6** |
| ***Resource Dilution/Confluence Model*** | | | | | | |
| Taking Leading Roles in Publication | 0.080*** | | | 0.051*** | | |
| | (0.005) | | | (0.005) | | |
| Advisor Stimulation | 0.033*** | | | -0.011*** | | |
| | (0.005) | | | (0.002) | | |
| Sibling Stimulation | 0.016 | | | 0.000 | | |
| | (0.009) | | | (0.002) | | |
| ***Role Specialization*** | | | | | | |
| Knowledge Breadth | | 0.023** | | | -0.002 | |
| | | (0.007) | | | (0.004) | |
| Similarity with Cohorts | | 0.172 | | | 0.111 | |
| | | (0.160) | | | (0.068) | |
| Distance from Advisor | | -1.542*** | | | -0.815*** | |
| | | (0.288) | | | (0.162) | |
| Distance from Advisor (Squared) | | 1.239*** | | | 0.578*** | |
| | | (0.293) | | | (0.159) | |
| ***Family Dynamics*** | | | | | | |
| Research Network Expansion | | | 0.012 | | | 0.010 |
| | | | (0.042) | | | (0.011) |
| Funding Expansion | | | -0.002 | | | -0.001 |
| | | | (0.003) | | | (0.001) |
| ***Author Race (Ref: White)*** | | | | | | |
| Asian/Pacific Islander (API) | 0.456*** | 0.865*** | 0.914*** | 0.331*** | 0.990*** | 0.962*** |
| | (0.042) | (0.017) | (0.047) | (0.033) | (0.014) | (0.025) |
| Black | 0.136 | -0.256** | -0.269 | -0.254* | -0.180*** | -0.155*** |
| | (0.230) | (0.095) | (0.312) | (0.106) | (0.017) | (0.041) |
| Hispanic | -0.048 | 0.014 | -0.207 | -0.163** | -0.119*** | -0.117*** |
| | (0.098) | (0.036) | (0.121) | (0.051) | (0.010) | (0.020) |
| AIAN | 1.133* | 0.031 | -0.240 | 0.319 | -0.005 | 0.243** |
| | (0.508) | (0.452) | (1.284) | (0.353) | (0.078) | (0.077) |
| Other/Unidentifiable | -0.165 | 0.018 | 0.092 | -0.288** | -0.029 | -0.098* |
| | (0.223) | (0.079) | (0.240) | (0.104) | (0.020) | (0.043) |
| ***Author Gender (Ref: Male)*** | | | | | | |
| Female | 0.043 | -0.111$^{***}$ | -0.141$^{**}$ | -0.341$^{***}$ | -0.179$^{***}$ | -0.167$^{***}$ |
| | (0.041) | (0.015) | (0.046) | (0.030) | (0.009) | (0.019) |
| Other/Unidentifiable | -0.230 | -0.759$^{***}$ | -0.489$^{**}$ | -0.529$^{***}$ | -0.895$^{***}$ | -0.920$^{***}$ |
| | (0.167) | (0.057) | (0.178) | (0.082) | (0.017) | (0.042) |
| Log(Mentor Experience) | -0.023 | -0.088$^{***}$ | -0.103 | -0.007 | -0.009 | -0.010 |
| | (0.050) | (0.019) | (0.067) | (0.040) | (0.011) | (0.028) |
| Log(Previous Publications) | 0.033 | 0.645$^{***}$ | 0.465$^{***}$ | 0.780$^{***}$ | 0.675$^{***}$ | 0.185$^{**}$ |
| | (0.053) | (0.036) | (0.085) | (0.066) | (0.055) | (0.067) |
| Constant | | | | 0.522$^{***}$ | 0.609$^{***}$ | 0.400$^{***}$ |
| | | | | (0.087) | (0.046) | (0.062) |
| Family/Year/Field Fixed Effects | YES | YES | YES | YES | YES | YES |
| Adjusted $R^2$ | | | | 0.043 | 0.023 | 0.028 |
| *N* | 202862 | 1179153 | 301427 | 139817 | 1201327 | 286812 |

*Note:* $^{*}$ *$p < 0.05$,* $^{**}$ *$p < 0.01$,* $^{***}$ *$p < 0.001$. All tests are two-tailed.*

Table A5_2. Effect Pathways from Intellectual Birth Order to Career Outcomes (Publication)

| | **Log(Short-Term Pubs)** | | | **Log(Long-Term Pubs)** | | |
|---|---|---|---|---|---|---|
| | **Model 1** | **Model 2** | **Model 3** | **Model 4** | **Model 5** | **Model 6** |
| ***Resource Dilution/Confluence Model*** | | | | | | |
| Taking Leading Roles in Publication | 0.149$^{***}$ | | | 0.141$^{***}$ | | |
| | (0.002) | | | (0.002) | | |
| Advisor Stimulation | 0.013$^{***}$ | | | 0.009$^{***}$ | | |
| | (0.003) | | | (0.002) | | |
| Sibling Stimulation | 0.004$^{***}$ | | | 0.005$^{***}$ | | |
| | (0.001) | | | (0.001) | | |
| ***Role Specialization*** | | | | | | |
| Knowledge Breadth | | 0.004$^{***}$ | | | 0.004$^{***}$ | |
| | | (0.001) | | | (0.001) | |
| Similarity with Cohorts | | 0.034 | | | 0.017 | |
| | | (0.021) | | | (0.023) | |
| Distance from Advisor | | -0.191$^{***}$ | | | -0.211$^{***}$ | |
| | | (0.036) | | | (0.041) | |
| Distance from Advisor (Squared) | | 0.119$^{***}$ | | | 0.130$^{**}$ | |
| | | (0.035) | | | (0.040) | |
| ***Family Dynamics*** | | | | | | |
| Research Network Expansion | | | 0.005 | | | 0.011 |
| | | | (0.006) | | | (0.007) |
| Funding Expansion | | | 0.000 | | | 0.001 |
| | | | (0.000) | | | (0.000) |
| ***Author Race (Ref: White)*** | | | | | | |
| Asian/Pacific Islander (API) | 0.022+ | 0.042$^{***}$ | 0.060$^{***}$ | 0.140$^{***}$ | 0.072$^{***}$ | 0.103$^{***}$ |
| | (0.012) | (0.003) | (0.009) | (0.014) | (0.003) | (0.010) |
| Black | -0.130$^{*}$ | -0.040$^{***}$ | -0.025 | -0.085 | -0.044$^{***}$ | -0.026 |
| | (0.057) | (0.008) | (0.035) | (0.072) | (0.009) | (0.036) |
| Hispanic | -0.112$^{***}$ | -0.010$^{*}$ | -0.030+ | -0.117$^{***}$ | -0.000 | -0.025 |
| | (0.026) | (0.005) | (0.017) | (0.031) | (0.005) | (0.019) |
| AIAN | 2.442$^{***}$ | 0.199 | 1.015 | 1.485$^{***}$ | 0.150 | 1.320 |
| | (0.286) | (0.161) | (0.667) | (0.430) | (0.179) | (0.838) |
| Other/Unidentifiable | 0.011 | -0.024$^{**}$ | -0.003 | 0.036 | -0.017 | 0.008 |
| | (0.051) | (0.008) | (0.033) | (0.061) | (0.009) | (0.037) |
| ***Author Gender (Ref: Male)*** | | | | | | |
| Female | -0.065$^{***}$ | -0.044$^{***}$ | -0.046$^{***}$ | -0.063$^{***}$ | -0.048$^{***}$ | -0.044$^{***}$ |
| | (0.010) | (0.002) | (0.007) | (0.013) | (0.002) | (0.008) |
| Other/Unidentifiable | -0.102$^{*}$ | -0.167$^{***}$ | -0.169$^{***}$ | -0.139$^{**}$ | -0.196$^{***}$ | -0.202$^{***}$ |
| | (0.040) | (0.005) | (0.022) | (0.049) | (0.006) | (0.025) |
| Log(Mentor Experience) | -0.004 | -0.006$^{*}$ | 0.014 | 0.008 | -0.004 | 0.005 |
| | (0.013) | (0.003) | (0.011) | (0.015) | (0.003) | (0.013) |
| Log(Previous Publications) | 0.304$^{***}$ | 1.770$^{***}$ | 1.805$^{***}$ | 0.474$^{***}$ | 1.978$^{***}$ | 1.995$^{***}$ |
| | (0.014) | (0.011) | (0.028) | (0.016) | (0.013) | (0.033) |
| Constant | 1.239$^{***}$ | 0.320$^{***}$ | 0.288$^{***}$ | 1.326$^{***}$ | 0.342$^{***}$ | 0.297$^{***}$ |
| | (0.026) | (0.011) | (0.025) | (0.031) | (0.012) | (0.028) |
| Family/Year/Field Fixed Effects | YES | YES | YES | YES | YES | YES |
| Adjusted $R^2$ | 0.365 | 0.169 | 0.194 | 0.285 | 0.154 | 0.180 |
| *N* | 76947 | 799218 | 76044 | 76947 | 799218 | 76044 |

*Note:* $^{*}$ *$p < 0.05$,* $^{**}$ *$p < 0.01$,* $^{***}$ *$p < 0.001$. All tests are two-tailed.*

Table A6_1. Cross-Sample Heterogeneity: Student Gender (Male vs. Female)

| | **Model 1** | **Model 2** | **Model 3** | **Model 4** | **Model 5** | **Model 6** | **Model 7** | **Model 8** |
|---|---|---|---|---|---|---|---|---|
| | Transition to Advisorship | | Students Count | | Short Term Pub | | Long Term Pub | |
| | Male | Female | Male | Female | Male | Female | Male | Female |
| ***Intellectual Birth Order (Ref: First Student)*** | | | | | | | | |
| 2-5 | -0.082$^{***}$ | -0.048 | 0.001 | -0.005 | -0.010$^{*}$ | 0.002 | -0.008 | 0.001 |
| | (0.022) | (0.037) | (0.026) | (0.016) | (0.005) | (0.005) | (0.005) | (0.007) |
| 6-10 | -0.136$^{***}$ | -0.138$^{**}$ | -0.023 | -0.055$^{**}$ | -0.013+ | 0.001 | -0.012 | -0.001 |
| | (0.033) | (0.052) | (0.038) | (0.020) | (0.007) | (0.007) | (0.008) | (0.009) |
| 11-15 | -0.212$^{***}$ | -0.138$^{*}$ | -0.074 | -0.078$^{**}$ | -0.015 | -0.003 | -0.013 | -0.004 |
| | (0.037) | (0.067) | (0.046) | (0.028) | (0.010) | (0.010) | (0.011) | (0.011) |
| 16-20 | -0.230$^{***}$ | -0.176$^{*}$ | -0.135$^{*}$ | -0.102$^{**}$ | -0.024$^{*}$ | -0.010 | -0.028$^{*}$ | -0.010 |
| | (0.058) | (0.076) | (0.054) | (0.033) | (0.011) | (0.009) | (0.012) | (0.010) |
| 21-30 | -0.294$^{***}$ | -0.142+ | -0.142$^{*}$ | -0.088$^{*}$ | -0.027$^{*}$ | -0.006 | -0.025+ | -0.012 |
| | (0.057) | (0.078) | (0.061) | (0.039) | (0.013) | (0.011) | (0.014) | (0.011) |
| ≥31 | -0.247$^{***}$ | -0.250$^{*}$ | -0.178$^{*}$ | -0.126$^{*}$ | -0.020 | -0.019 | -0.015 | -0.025+ |
| | (0.068) | (0.110) | (0.078) | (0.053) | (0.014) | (0.014) | (0.014) | (0.013) |
| ***Author Race (Ref: White)*** | | | | | | | | |
| Asian/Pacific Islander (API) | 0.844$^{***}$ | 0.935$^{***}$ | 1.012$^{***}$ | 0.996$^{***}$ | 0.044$^{***}$ | 0.061$^{***}$ | 0.076$^{***}$ | 0.089$^{***}$ |
| | (0.039) | (0.044) | (0.049) | (0.037) | (0.007) | (0.005) | (0.005) | (0.004) |
| Black | -0.276$^{*}$ | -0.276 | -0.292$^{***}$ | -0.070$^{***}$ | -0.047$^{***}$ | -0.040$^{**}$ | -0.051$^{***}$ | -0.049$^{*}$ |
| | (0.111) | (0.189) | (0.038) | (0.018) | (0.009) | (0.014) | (0.013) | (0.019) |
| Hispanic | -0.036 | 0.136$^{*}$ | -0.187$^{***}$ | -0.024 | -0.030$^{***}$ | 0.019$^{*}$ | -0.019$^{*}$ | 0.022+ |
| | (0.047) | (0.058) | (0.024) | (0.014) | (0.009) | (0.009) | (0.008) | (0.012) |
| AIAN | -24.728$^{***}$ | 0.177 | -0.115 | -0.120 | 0.284 | 0.075 | 0.220 | 0.031 |
| | (0.813) | (0.698) | (0.256) | (0.179) | (0.275) | (0.218) | (0.326) | (0.246) |
| Other/Unidentifiable | 0.012 | -0.018 | -0.195$^{***}$ | -0.001 | -0.038$^{**}$ | -0.025 | -0.028$^{*}$ | -0.028 |
| | (0.072) | (0.137) | (0.052) | (0.027) | (0.011) | (0.015) | (0.013) | (0.019) |
| ***Mentor/Mentee Abilities*** | | | | | | | | |
| Log(Mentor Experience) | -0.023 | -0.116$^{**}$ | -0.020 | 0.007 | 0.001 | -0.006 | 0.000 | -0.005 |
| | (0.027) | (0.039) | (0.019) | (0.012) | (0.004) | (0.004) | (0.004) | (0.006) |
| Log(Previous Publications) | 0.665$^{***}$ | 0.614$^{***}$ | 0.910$^{***}$ | 0.191$^{*}$ | 1.768$^{***}$ | 1.729$^{***}$ | 1.968$^{***}$ | 1.964$^{***}$ |
| | (0.034) | (0.067) | (0.127) | (0.082) | (0.028) | (0.028) | (0.032) | (0.033) |
| Constant | | | 0.468$^{***}$ | 0.198$^{***}$ | 0.300$^{***}$ | 0.197$^{***}$ | 0.314$^{***}$ | 0.209$^{***}$ |
| | | | (0.041) | (0.028) | (0.008) | (0.006) | (0.008) | (0.007) |
| Intellectual Family Fixed Effects | YES | YES | YES | YES | YES | YES | YES | YES |
| Year Fixed Effects | YES | YES | YES | YES | YES | YES | YES | YES |
| Field Fixed Effects | YES | YES | YES | YES | YES | YES | YES | YES |
| *N* | 707364 | 555061 | 726930 | 533742 | 495149 | 341669 | 495149 | 341669 |

*Note: $^{*}$ p < 0.05, $^{**}$ p < 0.01, $^{***}$ p < 0.001. All tests are two-tailed.*

Table A6_2. Cross-Sample Heterogeneity: Student Race (White vs. Others)

| | **Model 1** | **Model 2** | **Model 3** | **Model 4** | **Model 5** | **Model 6** | **Model 7** | **Model 8** |
|---|---|---|---|---|---|---|---|---|
| | Transition to Advisorship | | Students Count | | Short Term Pub | | Long Term Pub | |
| | White | Others | White | Others | White | Others | White | Others |
| ***Intellectual Birth Order (Ref: First Student)*** | | | | | | | | |
| 2-5 | -0.111$^{***}$ | -0.039 | -0.019 | 0.015 | -0.009$^{**}$ | 0.006 | -0.008 | 0.015 |
| | (0.023) | (0.039) | (0.014) | (0.044) | (0.003) | (0.010) | (0.004) | (0.011) |
| 6-10 | -0.183$^{***}$ | -0.065 | -0.057$^{**}$ | -0.013 | -0.013$^{*}$ | 0.006 | -0.012 | 0.013 |
| | (0.033) | (0.055) | (0.021) | (0.055) | (0.005) | (0.013) | (0.007) | (0.014) |
| 11-15 | -0.224$^{***}$ | -0.124$^{*}$ | -0.085$^{***}$ | -0.098 | -0.016$^{*}$ | 0.009 | -0.013 | 0.019 |
| | (0.045) | (0.062) | (0.023) | (0.078) | (0.007) | (0.018) | (0.008) | (0.019) |
| 16-20 | -0.226$^{***}$ | -0.140 | -0.113$^{***}$ | -0.166 | -0.024$^{***}$ | -0.011 | -0.019$^{*}$ | -0.013 |
| | (0.053) | (0.081) | (0.028) | (0.084) | (0.007) | (0.017) | (0.008) | (0.020) |
| 21-30 | -0.381$^{***}$ | -0.036 | -0.173$^{***}$ | -0.049 | -0.025$^{*}$ | -0.006 | -0.023$^{*}$ | -0.003 |
| | (0.058) | (0.091) | (0.031) | (0.113) | (0.010) | (0.024) | (0.011) | (0.025) |
| ≥31 | -0.297$^{***}$ | -0.073 | -0.210$^{***}$ | -0.103 | -0.022$^{*}$ | -0.023 | -0.013 | -0.032 |
| | (0.086) | (0.098) | (0.040) | (0.119) | (0.010) | (0.033) | (0.012) | (0.033) |
| ***Author Gender (Ref: Male)*** | | | | | | | | |
| Female | -0.199$^{***}$ | 0.028 | -0.257$^{***}$ | 0.039 | -0.046$^{***}$ | -0.040$^{***}$ | -0.052$^{***}$ | -0.041$^{***}$ |
| | (0.024) | (0.023) | (0.019) | (0.033) | (0.002) | (0.004) | (0.003) | (0.006) |
| Other/Unidentifiable | -0.337$^{***}$ | -0.793$^{***}$ | -0.321$^{***}$ | -1.074$^{***}$ | -0.109$^{***}$ | -0.186$^{***}$ | -0.117$^{***}$ | -0.222$^{***}$ |
| | (0.078) | (0.076) | (0.041) | (0.047) | (0.008) | (0.010) | (0.010) | (0.010) |
| ***Mentor/Mentee Abilities*** | | | | | | | | |
| Log(Mentor Experience) | -0.016 | -0.089$^{*}$ | -0.017 | -0.022 | -0.001 | -0.013$^{*}$ | 0.000 | -0.016$^{**}$ |
| | (0.025) | (0.037) | (0.009) | (0.033) | (0.003) | (0.005) | (0.003) | (0.006) |
| Log(Previous Publications) | 0.589*** | 0.652*** | 0.367*** | 0.917*** | 1.601*** | 1.879*** | 1.752*** | 2.136*** |
| | (0.040) | (0.046) | (0.061) | (0.163) | (0.028) | (0.039) | (0.028) | (0.040) |
| Constant | | | 0.463*** | 1.309*** | 0.262*** | 0.366*** | 0.275*** | 0.405*** |
| | | | (0.021) | (0.077) | (0.005) | (0.015) | (0.005) | (0.014) |
| Intellectual Family Fixed Effects | YES | YES | YES | YES | YES | YES | YES | YES |
| Year Fixed Effects | YES | YES | YES | YES | YES | YES | YES | YES |
| Field Fixed Effects | YES | YES | YES | YES | YES | YES | YES | YES |
| *N* | 968517 | 317469 | 954328 | 354402 | 654253 | 221066 | 654253 | 221066 |

*Note:* * *p < 0.05,* ** *p < 0.01,* *** *p < 0.001. All tests are two-tailed.*

Table A6_3. Cross-Sample Heterogeneity: University Type (R1 vs. non-R1)

| | **Model 1** | **Model 2** | **Model 3** | **Model 4** | **Model 5** | **Model 6** | **Model 7** | **Model 8** |
|---|---|---|---|---|---|---|---|---|
| | Transition to Advisorship | | Students Count | | Short Term Pub | | Long Term Pub | |
| | R1 | non-R1 | R1 | non-R1 | R1 | non-R1 | R1 | non-R1 |
| ***Intellectual Birth Order (Ref: First Student)*** | | | | | | | | |
| 2-5 | -0.071*** | -0.085 | -0.002 | -0.001 | -0.008* | 0.002 | -0.005 | 0.004 |
| | (0.021) | (0.049) | (0.016) | (0.033) | (0.003) | (0.007) | (0.003) | (0.008) |
| 6-10 | -0.128*** | -0.141* | -0.041 | -0.012 | -0.011* | -0.001 | -0.010 | -0.001 |
| | (0.028) | (0.070) | (0.023) | (0.051) | (0.005) | (0.010) | (0.006) | (0.012) |
| 11-15 | -0.181*** | -0.203* | -0.074* | -0.106 | -0.015 | 0.004 | -0.012 | 0.004 |
| | (0.035) | (0.094) | (0.031) | (0.054) | (0.008) | (0.012) | (0.009) | (0.014) |
| 16-20 | -0.192** | -0.173 | -0.134*** | -0.090 | -0.025** | -0.012 | -0.025** | -0.007 |
| | (0.061) | (0.100) | (0.037) | (0.077) | (0.009) | (0.014) | (0.009) | (0.015) |
| 21-30 | -0.242*** | -0.150 | -0.125** | -0.128 | -0.026** | -0.003 | -0.026* | -0.004 |
| | (0.049) | (0.144) | (0.042) | (0.086) | (0.010) | (0.016) | (0.012) | (0.019) |
| ≥31 | -0.188** | -0.287 | -0.169** | -0.148 | -0.028* | -0.014 | -0.027* | -0.010 |
| | (0.065) | (0.173) | (0.050) | (0.084) | (0.012) | (0.016) | (0.012) | (0.021) |
| ***Author Race (Ref: White)*** | | | | | | | | |
| Asian/Pacific Islander (API) | 0.888*** | 0.900*** | 1.025*** | 0.849*** | 0.047*** | 0.034*** | 0.077*** | 0.060*** |
| | (0.042) | (0.052) | (0.045) | (0.044) | (0.005) | (0.009) | (0.003) | (0.008) |
| Black | -0.274* | -0.267 | -0.162*** | -0.243*** | -0.038*** | -0.059*** | -0.042*** | -0.061*** |
| | (0.126) | (0.148) | (0.024) | (0.026) | (0.008) | (0.011) | (0.009) | (0.014) |
| Hispanic | 0.013 | 0.014 | -0.103$^{***}$ | -0.195$^{***}$ | -0.012 | -0.000 | -0.003 | 0.004 |
| | (0.045) | (0.092) | (0.015) | (0.027) | (0.007) | (0.009) | (0.008) | (0.011) |
| AIAN | 0.936$^{**}$ | -33.037$^{***}$ | -0.004 | -0.216 | 0.230 | -0.162$^{*}$ | 0.205 | -0.189$^{*}$ |
| | (0.362) | (0.693) | (0.112) | (0.194) | (0.135) | (0.081) | (0.149) | (0.084) |
| Other/Unidentifiable | -0.017 | -0.087 | 0.001 | -0.151$^{***}$ | -0.022$^{*}$ | -0.022 | -0.012 | -0.026 |
| | (0.071) | (0.175) | (0.036) | (0.038) | (0.009) | (0.016) | (0.010) | (0.019) |
| ***Author Gender (Ref: Male)*** | | | | | | | | |
| Female | -0.091$^{***}$ | -0.231$^{***}$ | -0.146$^{***}$ | -0.278$^{***}$ | -0.045$^{***}$ | -0.046$^{***}$ | -0.049$^{***}$ | -0.049$^{***}$ |
| | (0.019) | (0.030) | (0.014) | (0.018) | (0.002) | (0.004) | (0.003) | (0.004) |
| Other/Unidentifiable | -0.754$^{***}$ | -0.510$^{***}$ | -0.914$^{***}$ | -0.679$^{***}$ | -0.171$^{***}$ | -0.149$^{***}$ | -0.205$^{***}$ | -0.169$^{***}$ |
| | (0.062) | (0.097) | (0.034) | (0.054) | (0.008) | (0.009) | (0.009) | (0.012) |
| ***Mentor/Mentee Abilities*** | | | | | | | | |
| Log(Mentor Experience) | -0.045$^{*}$ | -0.022 | -0.014 | -0.018 | -0.003 | -0.003 | -0.002 | -0.006 |
| | (0.021) | (0.049) | (0.013) | (0.018) | (0.003) | (0.005) | (0.003) | (0.005) |
| Log(Previous Publications) | 0.658$^{***}$ | 0.578$^{***}$ | 0.666$^{***}$ | 0.731$^{***}$ | 1.773$^{***}$ | 1.727$^{***}$ | 1.982$^{***}$ | 1.941$^{***}$ |
| | (0.033) | (0.080) | (0.106) | (0.152) | (0.025) | (0.026) | (0.027) | (0.035) |
| Constant | | | 0.417$^{***}$ | 0.527$^{***}$ | 0.287$^{***}$ | 0.248$^{***}$ | 0.299$^{***}$ | 0.267$^{***}$ |
| | | | (0.030) | (0.034) | (0.006) | (0.008) | (0.006) | (0.009) |
| Intellectual Family Fixed Effects | YES | YES | YES | YES | YES | YES | YES | YES |
| Year Fixed Effects | YES | YES | YES | YES | YES | YES | YES | YES |
| Field Fixed Effects | YES | YES | YES | YES | YES | YES | YES | YES |
| *N* | 963297 | 322689 | 1078605 | 321260 | 746263 | 193855 | 746263 | 193855 |

*Note:* $^{*}$ $p < 0.05$, $^{**}$ $p < 0.01$, $^{***}$ $p < 0.001$. *All tests are two-tailed.*

Table A6_4. Cross-Sample Heterogeneity: Disciplinary Domains (STEM vs. Other Majors)

| | **Model 1** | **Model 2** | **Model 3** | **Model 4** | **Model 5** | **Model 6** | **Model 7** | **Model 8** |
|---|---|---|---|---|---|---|---|---|
| | Transition to Advisorship | | Students Count | | Short Term Pub | | Long Term Pub | |
| | STEM | non-STEM | STEM | non-STEM | STEM | non-STEM | STEM | non-STEM |
| ***Intellectual Birth Order (Ref: First Student)*** | | | | | | | | |
| 2-5 | -0.105$^{***}$ | -0.006 | -0.015 | -0.008 | -0.007 | -0.001 | -0.003 | -0.002 |
| | (0.026) | (0.036) | (0.020) | (0.028) | (0.005) | (0.005) | (0.006) | (0.005) |
| 6-10 | -0.163$^{***}$ | -0.052 | -0.046 | -0.047 | -0.011 | -0.004 | -0.006 | -0.006 |
| | (0.036) | (0.054) | (0.030) | (0.037) | (0.007) | (0.007) | (0.009) | (0.008) |
| 11-15 | -0.211*** | -0.064 | -0.084* | -0.105* | -0.010 | -0.010 | -0.002 | -0.015 |
| | (0.043) | (0.063) | (0.036) | (0.045) | (0.010) | (0.006) | (0.012) | (0.008) |
| 16-20 | -0.204** | -0.118 | -0.149** | -0.122* | -0.023 | -0.019* | -0.021 | -0.019 |
| | (0.067) | (0.077) | (0.047) | (0.050) | (0.014) | (0.008) | (0.015) | (0.011) |
| 21-30 | -0.275*** | -0.040 | -0.144* | -0.117* | -0.027 | -0.014 | -0.019 | -0.022 |
| | (0.064) | (0.104) | (0.057) | (0.053) | (0.014) | (0.011) | (0.017) | (0.012) |
| ≥31 | -0.236*** | -0.059 | -0.227** | -0.092 | -0.025 | -0.028** | -0.016 | -0.032* |
| | (0.064) | (0.112) | (0.067) | (0.054) | (0.015) | (0.009) | (0.016) | (0.013) |
| ***Author Race (Ref: White)*** | | | | | | | | |
| Asian/Pacific Islander (API) | 0.956*** | 0.678*** | 1.107*** | 0.658*** | 0.052*** | 0.022** | 0.086*** | 0.041*** |
| | (0.040) | (0.046) | (0.048) | (0.041) | (0.005) | (0.007) | (0.004) | (0.006) |
| Black | -0.281* | -0.166 | -0.154*** | -0.236*** | -0.044*** | -0.046*** | -0.053*** | -0.051*** |
| | (0.118) | (0.161) | (0.024) | (0.021) | (0.010) | (0.010) | (0.012) | (0.013) |
| Hispanic | -0.066 | 0.127 | -0.128*** | -0.106*** | -0.023** | 0.005 | -0.012 | 0.010 |
| | (0.045) | (0.066) | (0.017) | (0.020) | (0.008) | (0.007) | (0.008) | (0.008) |
| AIAN | -0.019 | 0.790 | 0.002 | -0.129 | 0.512** | -0.203* | 0.469* | -0.208* |
| | (0.731) | (0.843) | (0.201) | (0.169) | (0.164) | (0.097) | (0.193) | (0.098) |
| Other/Unidentifiable | -0.071 | 0.006 | 0.008 | -0.109* | -0.020 | -0.033* | -0.007 | -0.033* |
| | (0.083) | (0.103) | (0.040) | (0.042) | (0.011) | (0.015) | (0.015) | (0.016) |
| ***Author Gender (Ref: Male)*** | | | | | | | | |
| Female | -0.065*** | -0.166*** | -0.117*** | -0.253*** | -0.051*** | -0.037*** | -0.055*** | -0.041*** |
| | (0.017) | (0.030) | (0.016) | (0.019) | (0.003) | (0.002) | (0.003) | (0.003) |
| Other/Unidentifiable | -0.642*** | -0.857*** | -0.908*** | -0.724*** | -0.177*** | -0.141*** | -0.209*** | -0.170*** |
| | (0.063) | (0.109) | (0.037) | (0.038) | (0.009) | (0.009) | (0.009) | (0.011) |
| ***Mentor/Mentee Abilities*** | | | | | | | | |
| Log(Mentor Experience) | -0.028 | -0.091* | -0.016 | -0.015 | -0.003 | -0.004 | -0.003 | -0.004 |
| | (0.017) | (0.042) | (0.016) | (0.016) | (0.003) | (0.004) | (0.004) | (0.005) |
| Log(Previous Publications) | 0.643*** | 0.691*** | 0.692*** | 0.600*** | 1.799*** | 1.622*** | 2.010*** | 1.820*** |
| | (0.035) | (0.071) | (0.123) | (0.114) | (0.026) | (0.040) | (0.032) | (0.035) |
| Constant | | | 0.433*** | 0.486*** | 0.314*** | 0.220*** | 0.322*** | 0.242*** |
| | | | (0.033) | (0.040) | (0.008) | (0.007) | (0.007) | (0.008) |
| Intellectual Family Fixed Effects | YES | YES | YES | YES | YES | YES | YES | YES |
| Year Fixed Effects | YES | YES | YES | YES | YES | YES | YES | YES |
| Field Fixed Effects | YES | YES | YES | YES | YES | YES | YES | YES |
| *N* | 813960 | 472026 | 861652 | 476764 | 567724 | 329246 | 567724 | 329246 |

*Note:* $^{*}$ $p < 0.05$, $^{**}$ $p < 0.01$, $^{***}$ $p < 0.001$. *All tests are two-tailed.*

## Appendix C. Robustness Model Estimation Using Alternative Measures

In this section, we replicate several variables using alternative measurements to test the robustness of our models. Specifically, for resource dilution theory models, we replace the absolute number of papers with within-lineage *z*-scores or proportion (Table A7_1); and for the confluence model estimations, we replace the absolute number of papers with their proportions among all Ph.D.-stage publications (Table A7_2), to account for disciplinary heterogeneity, individual-level productivity, and temporal dynamics. Finally, we conduct a set of robustness tests for Mechanism 3 (family dynamics). We replace the funding expansion and research network expansion variables with the average funding amount and average network size per paper, and re-estimate the models (Table A7_3).

These results, reported in Tables A7_1–A7_3, are broadly consistent with those presented in the main text. The sample size in Table A7 models is smaller than that in the main text because we include only those lineages in which students exhibit variation in the mechanism variables so *z*-scores can be calculated.

## Appendix D. Notes on Selection Bias across Intellectual Birth Order

While propensity score weighting and mechanism regressions allow us to filter out much of the selection effect and isolate treatment effects, we recognize an important disanalogy with biological families: in intellectual families, both students' and advisors' risk preferences may jointly determine when and in what order a student enters a lineage—and therefore the student's intellectual birth order itself. To probe this concern, we conduct two additional tests reported in Table A8, examining how students' demographic and academic backgrounds predict their birth order within a lineage. We find that male students, white students, and those with more publications prior to entering their Ph.D. are more likely to occupy early positions in a lineage.

Table A7_1. Mechanism 1 Robustness Test for Resource Dilution Theory: With *Z*-Scores

| | **DV: Leading Papers Z-Score** | | **DV: Pub Impact Factor Z-Score** | |
|---|---|---|---|---|
| | **Model 1** | **Model 2** | **Model 3** | **Model 4** |
| ***Intellectual Birth Order (Ref: First Student)*** | | | | |
| 2-5 | -0.015 | -0.014 | -0.010 | -0.011 |
| | (0.017) | (0.017) | (0.017) | (0.017) |
| 6-10 | -0.034 | -0.031+ | -0.020 | -0.021 |
| | (0.022) | (0.022) | (0.022) | (0.023) |
| 11-15 | -0.049+ | -0.045+ | -0.026 | -0.026 |
| | (0.026) | (0.027) | (0.026) | (0.027) |
| 16-20 | $-0.063^{*}$ | $-0.059^{*}$ | -0.020 | -0.021 |
| | (0.029) | (0.030) | (0.029) | (0.030) |
| 21-30 | $-0.070^{*}$ | -0.066+ | -0.033 | -0.034 |
| | (0.033) | (0.034) | (0.033) | (0.034) |
| ≥31 | $-0.121^{**}$ | $-0.116^{**}$ | -0.040 | -0.041 |
| | (0.039) | (0.040) | (0.040) | (0.040) |
| Cohort Size | | -0.001 | | 0.000 |
| | | (0.002) | | (0.001) |
| ***Author Race (Ref: White)*** | | | | |
| Asian/Pacific Islander (API) | -0.014 | -0.014 | $-0.026^{**}$ | $-0.026^{**}$ |
| | (0.008) | (0.008) | (0.008) | (0.008) |
| Black | $-0.135^{***}$ | $-0.135^{***}$ | $-0.131^{**}$ | $-0.131^{**}$ |
| | (0.039) | (0.039) | (0.040) | (0.040) |
| Hispanic | $-0.182^{***}$ | $-0.182^{***}$ | $-0.084^{***}$ | $-0.084^{***}$ |
| | (0.019) | (0.019) | (0.019) | (0.019) |
| AIAN | 0.029 | 0.030 | 0.315 | 0.315 |
| | (0.504) | (0.504) | (0.564) | (0.564) |
| Other/Unidentifiable | 0.050 | 0.050 | -0.056 | -0.056 |
| | (0.037) | (0.037) | (0.037) | (0.037) |
| ***Author Gender (Ref: Male)*** | | | | |
| Female | $-0.048^{***}$ | $-0.048^{***}$ | $-0.017^{*}$ | $-0.017^{*}$ |
| | (0.008) | (0.008) | (0.008) | (0.008) |
| Other/Unidentifiable | -0.025 | -0.025 | -0.021 | -0.021 |
| | (0.029) | (0.029) | (0.031) | (0.031) |
| Log(Mentor Experience) | $-0.044^{***}$ | $-0.043^{***}$ | 0.002 | 0.002 |
| | (0.011) | (0.011) | (0.011) | (0.011) |
| Log(Previous Publications) | $0.618^{***}$ | $0.618^{***}$ | $0.239^{***}$ | $0.239^{***}$ |
| | (0.008) | (0.008) | (0.008) | (0.008) |
| Constant | $0.076^{***}$ | $0.078^{***}$ | 0.002 | 0.001 |
| | (0.021) | (0.021) | (0.021) | (0.021) |
| Family/Year/Field Fixed Effects | YES | YES | YES | YES |
| Adjusted $R^2$ | 0.070 | 0.070 | 0.015 | 0.015 |
| *N* | 127602 | 127602 | 134227 | 134227 |

*Note:* $^{*}$ $p < 0.05$, $^{**}$ $p < 0.01$, $^{***}$ $p < 0.001$. *All tests are two-tailed.*

Table A7_2. Mechanism 1 Robustness Test for Confluence Model: With Proportion of Papers

| | **DV: Adult Stimulation (Prop. of Papers)** | | **DV: Sibling Stimulation (Prop. of Papers)** | | **DV: Mixed Stimulation (Prop. of Papers)** | |
|---|---|---|---|---|---|---|
| | **Model 1** | **Model 2** | **Model 3** | **Model 4** | **Model 5** | **Model 6** |
| ***Intellectual Birth Order (Ref: First Student)*** | | | | | | |
| 2-5 | 0.000 | 0.000 | $0.021^{***}$ | $0.018^{***}$ | 0.001 | 0.000 |
| | (0.001) | (0.001) | (0.003) | (0.003) | (0.000) | (0.000) |
| 6-10 | -0.001 | -0.000 | $0.034^{***}$ | $0.027^{***}$ | 0.001 | 0.000 |
| | (0.002) | (0.002) | (0.004) | (0.004) | (0.001) | (0.001) |
| 11-15 | -0.002 | -0.001 | $0.043^{***}$ | $0.035^{***}$ | 0.001 | 0.001 |
| | (0.002) | (0.002) | (0.004) | (0.004) | (0.001) | (0.001) |
| 16-20 | -0.002 | -0.002 | $0.046^{***}$ | $0.037^{***}$ | $0.002^{*}$ | 0.001 |
| | (0.003) | (0.003) | (0.005) | (0.005) | (0.001) | (0.001) |
| 21-30 | -0.003 | -0.002 | $0.054^{***}$ | $0.043^{***}$ | $0.002^{*}$ | 0.002 |
| | (0.003) | (0.003) | (0.006) | (0.006) | (0.001) | (0.001) |
| ≥31 | -0.003 | -0.003 | $0.065^{***}$ | $0.054^{***}$ | 0.002 | 0.001 |
| | (0.003) | (0.003) | (0.007) | (0.007) | (0.001) | (0.001) |
| Cohort Size | | -0.000 | | $0.003^{***}$ | | $0.000^{*}$ |
| | | (0.000) | | (0.000) | | (0.000) |
| ***Author Race (Ref: White)*** | | | | | | |
| Asian/Pacific Islander (API) | 0.001 | 0.001 | $-0.005^{**}$ | $-0.005^{**}$ | 0.000 | 0.000 |
| | (0.001) | (0.001) | (0.002) | (0.002) | (0.000) | (0.000) |
| Black | 0.001 | 0.001 | 0.009 | 0.009 | 0.000 | 0.000 |
| | (0.003) | (0.003) | (0.006) | (0.006) | (0.001) | (0.001) |
| Hispanic | -0.002 | -0.002 | -0.002 | -0.002 | $-0.001^{*}$ | $-0.001^{*}$ |
| | (0.001) | (0.001) | (0.003) | (0.003) | (0.000) | (0.000) |
| AIAN | -0.004 | -0.004 | 0.418 | 0.419 | -0.001 | -0.001 |
| | (0.002) | (0.002) | (0.352) | (0.349) | (0.001) | (0.001) |
| Other/Unidentifiable | 0.005 | 0.005+ | 0.001 | 0.001 | 0.001 | 0.001 |
| | (0.003) | (0.003) | (0.006) | (0.006) | (0.001) | (0.001) |
| ***Author Gender (Ref: Male)*** | | | | | | |
| Female | 0.001 | 0.001 | $0.006^{***}$ | $0.006^{***}$ | $0.001^{*}$ | $0.001^{*}$ |
| | (0.001) | (0.001) | (0.001) | (0.001) | (0.000) | (0.000) |
| Other/Unidentifiable | 0.001 | 0.001 | 0.007 | 0.007 | -0.001 | -0.001 |
| | (0.002) | (0.002) | (0.006) | (0.006) | (0.001) | (0.001) |
| Log(Mentor Experience) | -0.001 | -0.001 | $0.007^{***}$ | $0.006^{***}$ | 0.000 | 0.000 |
| | (0.001) | (0.001) | (0.001) | (0.001) | (0.000) | (0.000) |
| Log(Previous Publications) | $-0.002^{***}$ | $-0.002^{***}$ | $-0.011^{***}$ | $-0.011^{***}$ | $-0.000^{***}$ | $-0.000^{***}$ |
| | (0.000) | (0.000) | (0.001) | (0.001) | (0.000) | (0.000) |
| Constant | $0.020^{***}$ | $0.020^{***}$ | $0.014^{***}$ | $0.009^{*}$ | 0.000 | -0.000 |
| | (0.002) | (0.002) | (0.004) | (0.004) | (0.001) | (0.001) |
| Family/Year/Field Fixed Effects | YES | YES | YES | YES | YES | YES |
| Adjusted $R^2$ | 0.134 | 0.134 | 0.516 | 0.516 | 0.054 | 0.054 |
| $N$ | 139407 | 139407 | 139407 | 139407 | 139407 | 139407 |

*Note:* $^{*}$ $p < 0.05$, $^{**}$ $p < 0.01$, $^{***}$ $p < 0.001$. *All tests are two-tailed.*

Table A7_3. Mechanism 4 Robustness Test: With Raw *Research Network Size and Funding Size*

| | **DV: Average Research Network Size of Advisor During PhD Years** | | **DV: Average Funding Size of Advisor During PhD Years** | |
|---|---|---|---|---|
| | **Model 1** | **Model 2** | **Model 3** | **Model 4** |
| ***Intellectual Birth Order (Ref: First Student)*** | | | | |
| 2-5 | 0.189 | 0.150 | -0.002*** | -0.002*** |
| | (0.150) | (0.154) | (0.000) | (0.000) |
| 6-10 | -0.030 | -0.116 | -0.001* | -0.001** |
| | (0.225) | (0.248) | (0.000) | (0.000) |
| 11-15 | -0.940** | -1.048** | -0.000 | -0.001 |
| | (0.336) | (0.364) | (0.001) | (0.001) |
| 16-20 | -1.793*** | -1.919*** | -0.001 | -0.001 |
| | (0.373) | (0.404) | (0.001) | (0.001) |
| 21-30 | -1.432* | -1.567* | -0.001 | -0.002 |
| | (0.653) | (0.680) | (0.001) | (0.001) |
| ≥31 | -2.431* | -2.565* | -0.010*** | -0.010*** |
| | (1.159) | (1.103) | (0.002) | (0.002) |
| Cohort Size | | 0.037 | | 0.000 |
| | | (0.031) | | (0.000) |
| ***Author Race (Ref: White)*** | | | | |
| Asian/Pacific Islander (API) | 0.090 | 0.090 | -0.001** | -0.001** |
| | (0.076) | (0.076) | (0.000) | (0.000) |
| Black | -0.027 | -0.027 | -0.001 | -0.001 |
| | (0.233) | (0.233) | (0.001) | (0.001) |
| Hispanic | 0.126 | 0.125 | -0.000 | -0.000 |
| | (0.161) | (0.161) | (0.000) | (0.000) |
| AIAN | -0.364 | -0.357 | 0.010 | 0.010 |
| | (0.358) | (0.357) | (0.012) | (0.012) |
| Other/Unidentifiable | -0.455 | -0.455 | -0.001* | -0.001* |
| | (0.285) | (0.285) | (0.001) | (0.001) |
| ***Author Gender (Ref: Male)*** | | | | |
| Female | 0.060 | 0.060 | 0.001*** | 0.001*** |
| | (0.051) | (0.051) | (0.000) | (0.000) |
| Other/Unidentifiable | 0.182 | 0.182 | 0.000 | 0.000 |
| | (0.139) | (0.139) | (0.000) | (0.000) |
| Log(Mentor Experience) | -0.332 | -0.353 | 0.001** | 0.001* |
| | (0.198) | (0.203) | (0.000) | (0.000) |
| Log(Previous Publications) | 1.994*** | 1.993*** | -0.000 | -0.000 |
| | (0.533) | (0.533) | (0.000) | (0.000) |
| Constant | 6.275*** | 6.182*** | 0.026*** | 0.026*** |
| | (0.322) | (0.309) | (0.001) | (0.001) |
| Family/Year/Field Fixed Effects | YES | YES | YES | YES |
| Adjusted $R^2$ | 0.737 | 0.737 | 0.643 | 0.643 |
| *N* | 1261809 | 1261809 | 1261809 | 1261809 |

*Note: * $p < 0.05$, ** $p < 0.01$, *** $p < 0.001$. All tests are two-tailed.*

Table A8. Using Students' Background to Predict Their Intellectual Birth Order

| | Model 1 | Model 2 | Model 3 | Model 4 | Model 5 |
|---|---|---|---|---|---|
| Log(Previous Publications) | -0.001 | | | | -0.010 |
| | (0.023) | | | | (0.023) |
| Dummy: Previous Collaboration | | -0.007 | | | -0.003 |
| | | (0.089) | | | (0.090) |
| ***Author Race (Ref: White)*** | | | | | |
| Asian/Pacific Islander (API) | | | $0.125^{***}$ | | $0.128^{***}$ |
| | | | (0.013) | | (0.013) |
| Black | | | 0.087 | | 0.089 |
| | | | (0.056) | | (0.056) |
| Hispanic | | | $0.101^{***}$ | | $0.102^{***}$ |
| | | | (0.025) | | (0.025) |
| AIAN | | | 0.420 | | 0.419 |
| | | | (0.601) | | (0.601) |
| Other/Unidentifiable | | | 0.036 | | 0.044 |
| | | | (0.041) | | (0.041) |
| ***Author Gender (Ref: Male)*** | | | | | |
| Female | | | | $0.020^{*}$ | $0.019^{*}$ |
| | | | | (0.010) | (0.010) |
| Other/Unidentifiable | | | | -0.026 | -0.060 |
| | | | | (0.041) | (0.042) |
| Constant | $7.337^{***}$ | $7.337^{***}$ | $7.302^{***}$ | $7.329^{***}$ | $7.295^{***}$ |
| | (0.000) | (0.000) | (0.004) | (0.004) | (0.006) |
| Family/Year/Field Fixed Effects | YES | YES | YES | YES | YES |
| Adjusted $R^2$ | 0.784 | 0.784 | 0.784 | 0.784 | 0.784 |
| *N* | 1399864 | 1399864 | 1399864 | 1399864 | 1399864 |

*Note:* $^{*}$ *$p < 0.05$,* $^{**}$ *$p < 0.01$,* $^{***}$ *$p < 0.001$. All tests are two-tailed.*

## Appendix E. Validation Survey through Prolific

To test the robustness of our measures and results, we recruited 167 Ph.D. degree holders through Prolific. The composition of this sample was designed to closely reflect the demographic distribution of students in our ProQuest dataset, with approximately 57% male, 43% female, 71% White, and 29% from other racial groups. To ensure data quality, we included only participants who spent more than 300 seconds (i.e., 5 minutes) completing the survey.

We first collected participants' basic demographic information, including age, gender, race, nationality, and the country where they completed their Ph.D. degree. Because participants were recruited globally, this sample also provides a robustness check beyond our observational data (i.e., ProQuest), which primarily focuses on U.S. academia.

Among the 167 valid responses collected through Prolific, 35.33% indicated that their advisor had students who dropped out, and 55.69% reported cases of students graduating earlier or later than expected. Although such occurrences are relatively common, only 7.19% involved any change to a respondent's own graduation time, and only 17.37% reported any change to a student's intellectual birth order. Because our predictor is a categorical variable that groups similar values into bins, we expect such disturbances to have an even smaller effect. Therefore, our operationalization of the birth order is solid.

We then asked a set of questions to assess whether the graduation years recorded in ProQuest can accurately represent students' intellectual birth order. Specifically, we asked the following question:

***B1. To the best of your knowledge, before you completed your PhD, did any students under your advisor withdraw from the program before completing their degree (drop out)?***

*B1_1. If yes, did any of these students leaving the program affect your expected PhD completion timing?*

*B1_2. If yes, did any students leaving the program change your position within your advisor's group? (For example, if you were originally your advisor's third student based on entry time, and became the second after someone left.)*

***B2. To the best of your knowledge, before you completed your PhD, did any students under your advisor complete their PhD earlier or later than expected?***

*B2_1. If some of your advisor's other students graduated earlier or later than expected, did this*

*affect your own expected PhD completion timing?*
*B2_2. If some of your advisor's other students graduated earlier or later than expected, did this change your position within your advisor's group? (For example, if you were originally your advisor's third student based on entry time, and became the fourth after someone graduated, based on graduation time.)*

Responses to these questions were framed as "Yes" or "No," and we calculated descriptive statistics based on participants' answers.

Next, we asked respondents to self-report their academic performance during their Ph.D. studies (C1) and to describe their first job after earning the Ph.D. degree (C2):
*C1. In terms of academic achievement, what percentile do you believe you fall into compared with other students of the same advisor, during your Ph.D. years?*

- *0-20% (worst performance) (1)*
- *20%-40% (2)*
- *40-60% (3)*
- *60%-80% (4)*
- *80%-100% (best performance) (5)*

*C2. *Immediately* After completing your PhD, which of the following best describes your professional position?*

- *Academic/research position at a university or research institute (tenure-track or equivalent) (1)*
- *Academic/research position at a university or research institute (non-tenure track) (2)*
- *Postdoctoral research position (3)*
- *Industry research position (e.g., corporate R&D, private lab) (4)*
- *Industry non-research position (e.g., management, consulting, policy) (5)*
- *Government or non-profit research position (6)*
- *Prefer not to answer (8)*
- *Other (please specify) (7)*

Although we do not find evidence that birth order influences the likelihood of Ph.D. graduates remaining in academia, we observe a downward gradient in students' self-assessed academic performance, as shown in Figure A1. This pattern, derived from survey data on a global sample of Ph.D. holders, complements the findings observed in our U.S.-based sample.

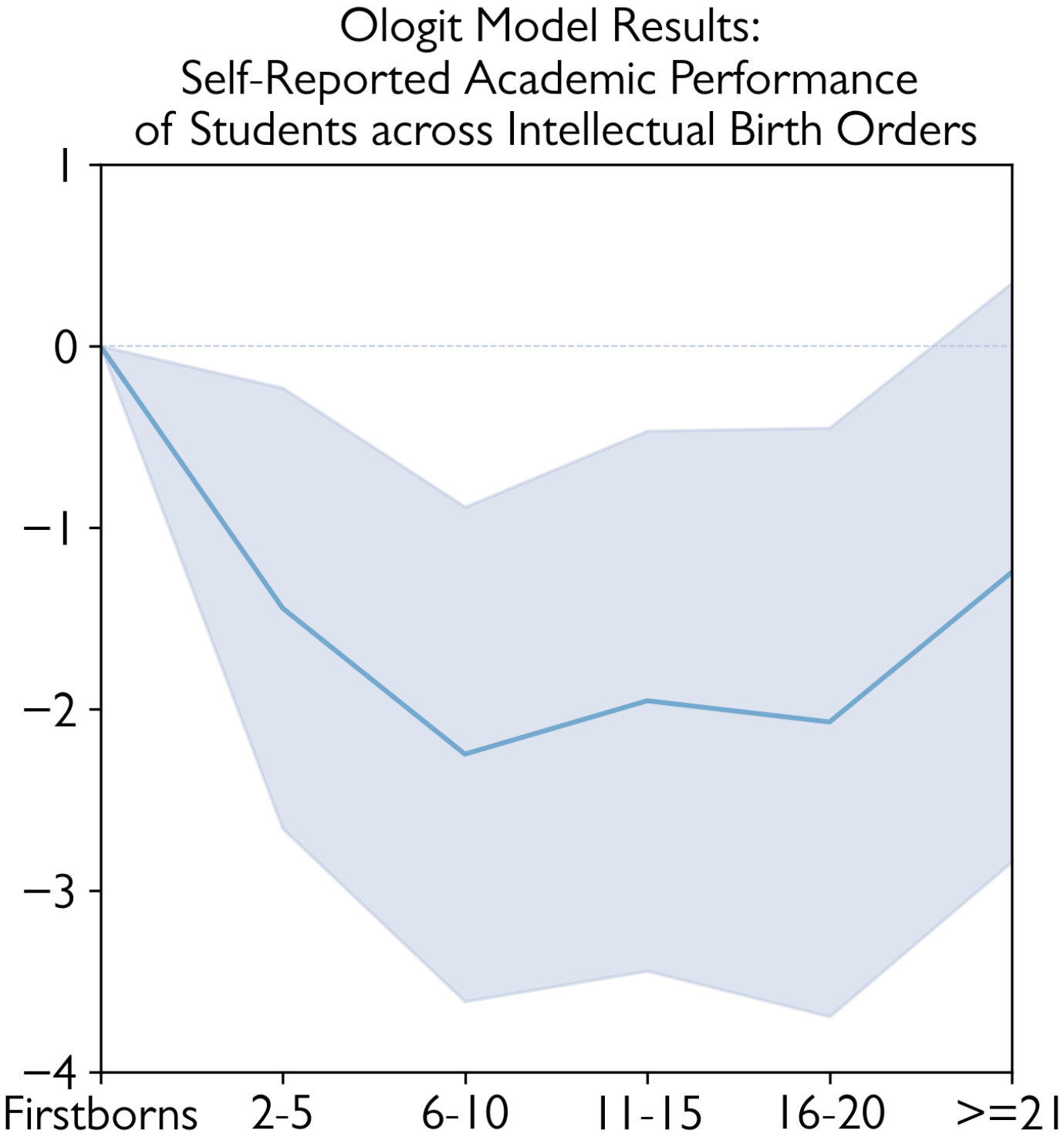

Figure A1. Intellectual Birth Order Effects Identified in an International, Survey-Based Dataset

(Note: Results are based on the Ologit command in Stata 18.0; all tests are two-tailed. Respondents' gender, race, discipline, Ph.D. program length, Ph.D. degree year, country of Ph.D. program, and size of lineage in 2025 are controlled.)

## Appendix F. Robustness Tests with Web of Science

Although OpenAlex is among the most widely used bibliometric data sources, recent studies have flagged a number of data-quality concerns (Zhou and Sun 2024). To probe the robustness of our findings, we re-matched the ProQuest Dissertations dataset to Web of Science (WoS)—a database known for stronger curation but narrower coverage, as it primarily indexes journal articles and excludes conference papers, preprints, and books.

We replicate the main models on this alternative dataset, excluding computer science—a field that relies heavily on conference papers poorly indexed in WoS. As shown in Table A9, the results are substantively consistent with our main findings.

Table A9. Robustness Tests with Web of Science Dataset

| | **DV: Transition to Advisorship (Cox with Strata)** | | **DV: Students Count (reghdfe)** | | **DV: Log(Short-Term Pubs) (reghdfe)** | | **DV: Log(Long-Term Pubs) (reghdfe)** | |
|---|---|---|---|---|---|---|---|---|
| | **Model 1** | **Model 2** | **Model 3** | **Model 4** | **Model 5** | **Model 6** | **Model 7** | **Model 8** |
| ***Intellectual Birth Order (Ref: First Student)*** | | | | | | | | |
| 2-5 | -0.011 | -0.073$^{**}$ | 0.005 | -0.007 | -0.009$^{***}$ | -0.002 | -0.010$^{***}$ | -0.003 |
| | (0.015) | (0.026) | (0.007) | (0.012) | (0.002) | (0.003) | (0.002) | (0.003) |
| 6-10 | -0.029 | -0.123$^{**}$ | 0.000 | -0.031+ | -0.021$^{***}$ | -0.005 | -0.023$^{***}$ | -0.007 |
| | (0.021) | (0.037) | (0.009) | (0.016) | (0.002) | (0.004) | (0.002) | (0.004) |
| 11-15 | -0.065$^{*}$ | -0.175$^{***}$ | -0.019+ | -0.068$^{***}$ | -0.028$^{***}$ | -0.008 | -0.031$^{***}$ | -0.010+ |
| | (0.028) | (0.047) | (0.011) | (0.020) | (0.003) | (0.005) | (0.003) | (0.005) |
| 16-20 | -0.054 | -0.169$^{**}$ | -0.025$^{*}$ | -0.092$^{***}$ | -0.040$^{***}$ | -0.014$^{*}$ | -0.042$^{***}$ | -0.017$^{**}$ |
| | (0.034) | (0.055) | (0.012) | (0.023) | (0.004) | (0.006) | (0.004) | (0.006) |
| 21-30 | -0.122$^{**}$ | -0.209$^{**}$ | -0.008 | -0.089$^{***}$ | -0.050$^{***}$ | -0.019$^{**}$ | -0.057$^{***}$ | -0.025$^{***}$ |
| | (0.041) | (0.066) | (0.014) | (0.027) | (0.004) | (0.007) | (0.005) | (0.007) |
| ≥31 | -0.122$^{*}$ | -0.195$^{*}$ | -0.023 | -0.136$^{***}$ | -0.069$^{***}$ | -0.026$^{**}$ | -0.067$^{***}$ | -0.023$^{*}$ |
| | (0.060) | (0.084) | (0.015) | (0.032) | (0.006) | (0.008) | (0.006) | (0.009) |
| ***Author Race (Ref: White)*** | | | | | | | | |
| Asian/Pacific Islander (API) | 1.051$^{***}$ | 0.909$^{***}$ | 0.859$^{***}$ | 0.791$^{***}$ | -0.101$^{***}$ | -0.123$^{***}$ | -0.091$^{***}$ | -0.110$^{***}$ |
| | (0.012) | (0.016) | (0.009) | (0.010) | (0.001) | (0.002) | (0.002) | (0.002) |
| Black | -0.488$^{***}$ | -0.277$^{**}$ | -0.054$^{***}$ | -0.052$^{***}$ | -0.025$^{***}$ | -0.019$^{**}$ | -0.046$^{***}$ | -0.036$^{***}$ |
| | (0.073) | (0.088) | (0.006) | (0.011) | (0.006) | (0.007) | (0.006) | (0.007) |
| Hispanic | 0.065$^{*}$ | 0.022 | -0.017$^{***}$ | -0.026$^{***}$ | 0.013$^{***}$ | 0.014$^{***}$ | 0.024$^{***}$ | 0.023$^{***}$ |
| | (0.029) | (0.035) | (0.004) | (0.007) | (0.003) | (0.004) | (0.004) | (0.004) |
| AIAN | 0.715 | 0.521 | 0.004 | 0.052 | -0.033 | 0.028 | -0.037 | 0.024 |
| | (0.561) | (0.651) | (0.035) | (0.054) | (0.075) | (0.075) | (0.100) | (0.113) |
| Other/Unidentifiable | -0.005 | -0.015 | 0.035$^{***}$ | 0.031$^{*}$ | -0.056$^{***}$ | -0.064$^{***}$ | -0.054$^{***}$ | -0.062$^{***}$ |
| | (0.065) | (0.077) | (0.010) | (0.015) | (0.005) | (0.006) | (0.006) | (0.007) |
| ***Author Gender (Ref: Male)*** | | | | | | | | |
| Female | -0.181$^{***}$ | -0.128$^{***}$ | -0.073$^{***}$ | -0.063$^{***}$ | -0.037$^{***}$ | -0.026$^{***}$ | -0.047$^{***}$ | -0.037$^{***}$ |
| | (0.011) | (0.014) | (0.004) | (0.006) | (0.001) | (0.001) | (0.001) | (0.002) |
| Other/Unidentifiable | -0.788$^{***}$ | -0.719$^{***}$ | -0.633$^{***}$ | -0.623$^{***}$ | -0.044$^{***}$ | -0.036$^{***}$ | -0.058$^{***}$ | -0.051$^{***}$ |
| | (0.042) | (0.051) | (0.011) | (0.015) | (0.003) | (0.003) | (0.003) | (0.004) |
| Log(Mentor Experience) | -0.010$^{*}$ | -0.027+ | -0.000 | -0.001 | 0.001$^{*}$ | -0.004$^{*}$ | 0.002$^{***}$ | -0.002 |
| | (0.005) | (0.015) | (0.002) | (0.006) | (0.001) | (0.002) | (0.001) | (0.002) |
| Log(Previous Publications) | 0.140$^{***}$ | 0.121$^{***}$ | -0.015$^{***}$ | -0.014$^{***}$ | 0.570$^{***}$ | 0.555$^{***}$ | 0.564$^{***}$ | 0.551$^{***}$ |
| | (0.013) | (0.017) | (0.003) | (0.004) | (0.003) | (0.003) | (0.003) | (0.004) |
| Constant | | | 0.202$^{***}$ | 0.247$^{***}$ | 0.251$^{***}$ | 0.254$^{***}$ | 0.246$^{***}$ | 0.247$^{***}$ |
| | | | (0.005) | (0.014) | (0.001) | (0.003) | (0.002) | (0.004) |
| Intellectual Family Fixed Effects | | YES | | YES | | YES | | YES |
| Year Fixed Effects | YES | YES | YES | YES | YES | YES | YES | YES |
| Field Fixed Effects | YES | YES | YES | YES | YES | YES | YES | YES |
| Adjusted $R^2$ | | | 0.024 | 0.020 | 0.136 | 0.162 | 0.109 | 0.130 |
| *N* | 1264561 | 1264561 | 1535728 | 1371998 | 1040233 | 919405 | 1040233 | 919405 |

*Note:* $^{*}$ $p < 0.05$, $^{**}$ $p < 0.01$, $^{***}$ $p < 0.001$. *All tests are two-tailed.*

## Appendix G. Notes on Causal Inference

In this section, we conduct two additional tests for causality.

We first employed a generalized propensity score weighting approach. Specifically, we estimated a generalized propensity score (GPS) for birth order, conditioned on a set of salient demographic covariates, such as gender and race, and characteristics that mentors typically consider when assessing a mentee's future potential, including publications achieved prior to graduate school and previous collaboration experience with the mentor. These factors largely determine whether a student joins a particular advisor and thus indirectly influence their rank within the intellectual family. We used an ordered logit model to estimate the GPS, as the treatment variable—birth order within intellectual families—is ordinal rather than binary (Robins et al. 2000; Naimi et al. 2014). From this model, we obtained the predicted probabilities ($\hat{p}$*), representing each individual's likelihood of belonging to each birth-order category.

Next, we constructed inverse probability weights. Each observation's realized birth-order category was matched with its corresponding predicted probability. The inverse probability weight (IPW) was then calculated as the reciprocal of the generalized propensity score (GPS), such that $\mathrm{IPW}_i = \frac{1}{P(\mathrm{Order}_i|X_i)}$.

This weighting scheme gives greater weight to observations that are less likely, given their covariates, to have their observed treatment level—thus balancing the covariate distribution across birth-order categories. To further stabilize weights and reduce variance, we computed stabilized inverse probability weights (SIPW) by multiplying the marginal probability of each treatment level ($p_{\mathrm{treat}}$) with the inverse of the GPS $\mathrm{SIPW}_i = \frac{P(\mathrm{Order} = j)}{P(\mathrm{Order}_i = j|X_i)}$. In the final stage, we applied these inverse probability weights in weighted fixed-effects regressions to estimate the effects of birth order on various scientific outcomes. All OLS models retained the same specification as the baseline analyses (as in Table 2), differing only by the inclusion of the weighting adjustment. Cox models (with transition to advisorship as the dependent variable) do not permit weight adjustment when high-dimensional strata are included. To address this limitation, we recode "transition to advisorship" as a binary outcome and estimate an OLS model with high-dimensional fixed effects and SIPW adjustment. While this model differs from the original specification, it offers insight into the existence and stability of the original results.

This propensity score weighting framework approximates a pseudo-population in which the distribution of observed covariates is balanced across birth-order categories. Under the assumption of conditional independence (i.e., the absence of unobserved confounding), this approach allows for a more credible estimation of the non-selective effects of intellectual birth order on subsequent career outcomes. The results, presented in Table A10, are broadly consistent with the baseline estimates reported in Table 2 in the main text. These results suggest that although some selection effects may exist, observed patterns remain consistent with a causal interpretation. Estimated effects likely approximate genuine treatment effects rather than mere selection bias.

Table A10. Robust Regression Estimates with Propensity Score Weighting

| | **Transition to Advisorship** | **Students Count** | **Log(Short-Term Pubs)** | **Log(Long-Term Pubs)** |
|---|---|---|---|---|
| | **Model 1** | **Model 2** | **Model 3** | **Model 4** |
| ***Intellectual Birth Order (Ref: First Student)*** | | | | |
| 2-5 | -0.002$^{*}$ | -0.004 | -0.006+ | -0.004 |
| | (0.001) | (0.017) | (0.003) | (0.004) |
| 6-10 | -0.005$^{***}$ | -0.039 | -0.009+ | -0.007 |
| | (0.001) | (0.023) | (0.005) | (0.006) |
| 11-15 | -0.008$^{***}$ | -0.084$^{**}$ | -0.011+ | -0.008 |
| | (0.002) | (0.029) | (0.006) | (0.007) |
| 16-20 | -0.009$^{***}$ | -0.128$^{***}$ | -0.022$^{**}$ | -0.021$^{**}$ |
| | (0.002) | (0.032) | (0.007) | (0.008) |
| 21-30 | -0.011$^{***}$ | -0.139$^{***}$ | -0.023$^{**}$ | -0.023$^{*}$ |
| | (0.002) | (0.037) | (0.008) | (0.009) |
| ≥31 | -0.007$^{**}$ | -0.176$^{***}$ | -0.023$^{*}$ | -0.022+ |
| | (0.003) | (0.045) | (0.010) | (0.011) |
| ***Author Race (Ref: White)*** | | | | |
| Asian/Pacific Islander (API) | 0.087$^{***}$ | 0.988$^{***}$ | 0.045$^{***}$ | 0.074$^{***}$ |
| | (0.001) | (0.013) | (0.002) | (0.003) |
| Black | -0.014$^{***}$ | -0.186$^{***}$ | -0.044$^{***}$ | -0.048$^{***}$ |
| | (0.001) | (0.016) | (0.008) | (0.009) |
| Hispanic | -0.005$^{***}$ | -0.124$^{***}$ | -0.010$^{*}$ | -0.002 |
| | (0.001) | (0.010) | (0.004) | (0.005) |
| AIAN | 0.026 | -0.014 | 0.192 | 0.160 |
| | (0.029) | (0.075) | (0.153) | (0.158) |
| Other/Unidentifiable | 0.004$^{*}$ | -0.036 | -0.024$^{**}$ | -0.015+ |
| | (0.002) | (0.019) | (0.008) | (0.009) |
| ***Author Gender (Ref: Male)*** | | | | |
| Female | -0.011$^{***}$ | -0.175$^{***}$ | -0.045$^{***}$ | -0.049$^{***}$ |
| | (0.000) | (0.008) | (0.002) | (0.002) |
| Other/Unidentifiable | -0.066$^{***}$ | -0.842$^{***}$ | -0.165$^{***}$ | -0.195$^{***}$ |
| | (0.001) | (0.017) | (0.004) | (0.005) |
| Log(Mentor Experience) | -0.002$^{*}$ | -0.015 | -0.003 | -0.002 |
| | (0.001) | (0.012) | (0.003) | (0.003) |
| Log(Previous Publications) | 0.061$^{***}$ | 0.695$^{***}$ | 1.764$^{***}$ | 1.972$^{***}$ |
| | (0.002) | (0.052) | (0.010) | (0.012) |
| Constant | 0.049$^{***}$ | 0.445$^{***}$ | 0.278$^{***}$ | 0.291$^{***}$ |
| | (0.001) | (0.020) | (0.004) | (0.005) |
| Intellectual Family Fixed Effects | YES | YES | YES | YES |
| Year Fixed Effects | YES | YES | YES | YES |
| Field Fixed Effects | YES | YES | YES | YES |
| Adjusted $R^2$ | 0.058 | 0.153 | 0.167 | 0.153 |
| *N* | 1399866 | 1399866 | 940119 | 940119 |

*Note:* $^{*}$ $p < 0.05$, $^{**}$ $p < 0.01$, $^{***}$ $p < 0.001$. *All tests are two-tailed.*

Next, we conduct a permutation test. Within each intellectual lineage, we randomly reshuffle birth order and re-estimate the same regression model in each iteration, recording the resulting coefficients and *t*-values. This process is repeated 500 times to generate empirical distributions for these statistics. We then compare the "true" estimates reported in Tables 2 with the distributions obtained through permutation. This non-parametric statistical test helps identify genuine structural relationships, by evaluating whether the observed effects exceed those expected under random assignment, thereby ruling out spurious correlations driven by sample composition or unobserved selection mechanisms.

The results of this permutation test are presented in Figure A2 (for coefficients) and Figure A3 (for *t*-values). In both cases, estimates from Table 2 (solid vertical lines) fall at the extreme tails of the permutation distributions, indicating that the estimated effects are highly unlikely to result from features of data structure.

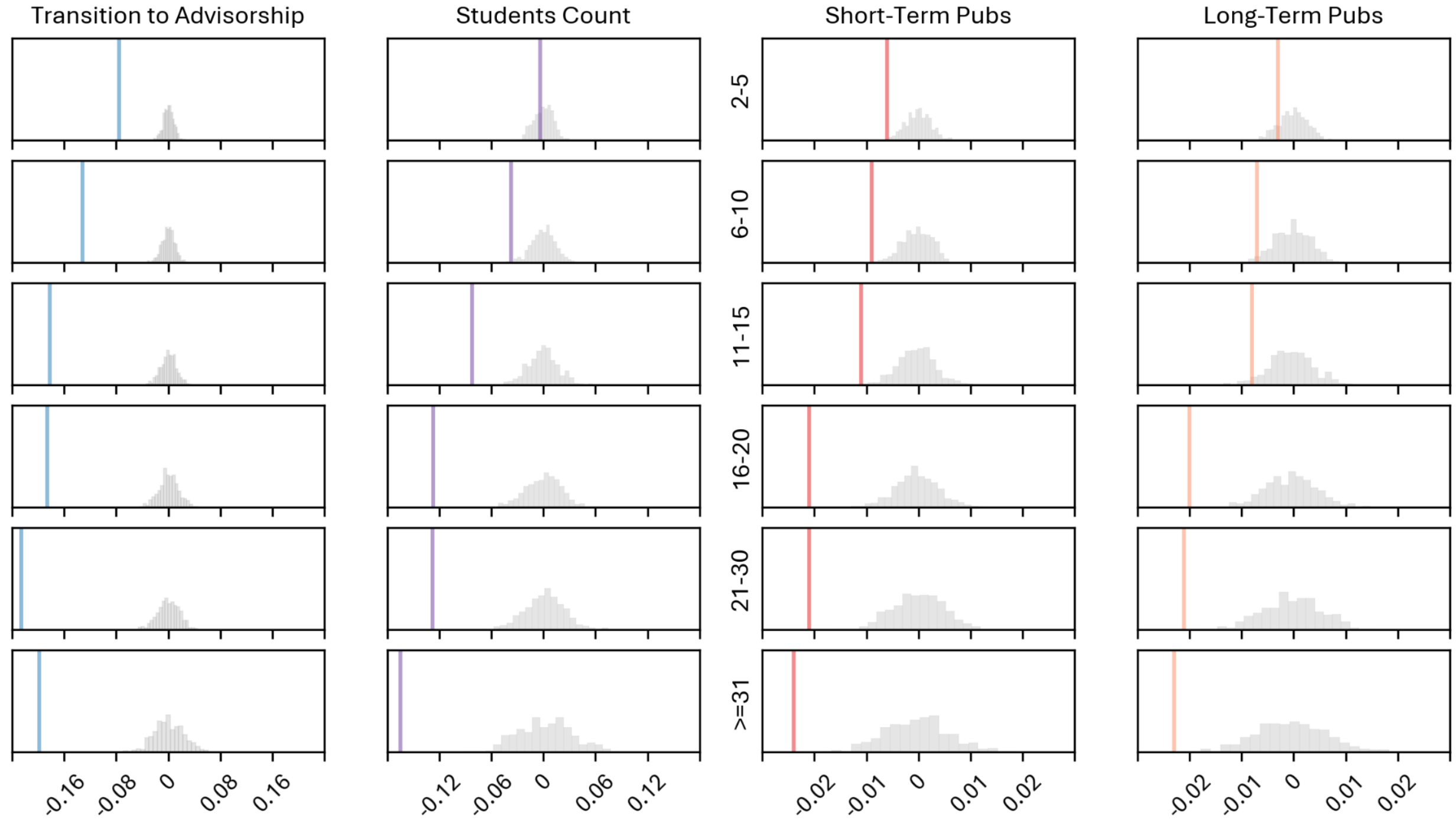

Figure A2. Permutation Test of Intellectual Birth-Order Effects (Coefficient Estimates)

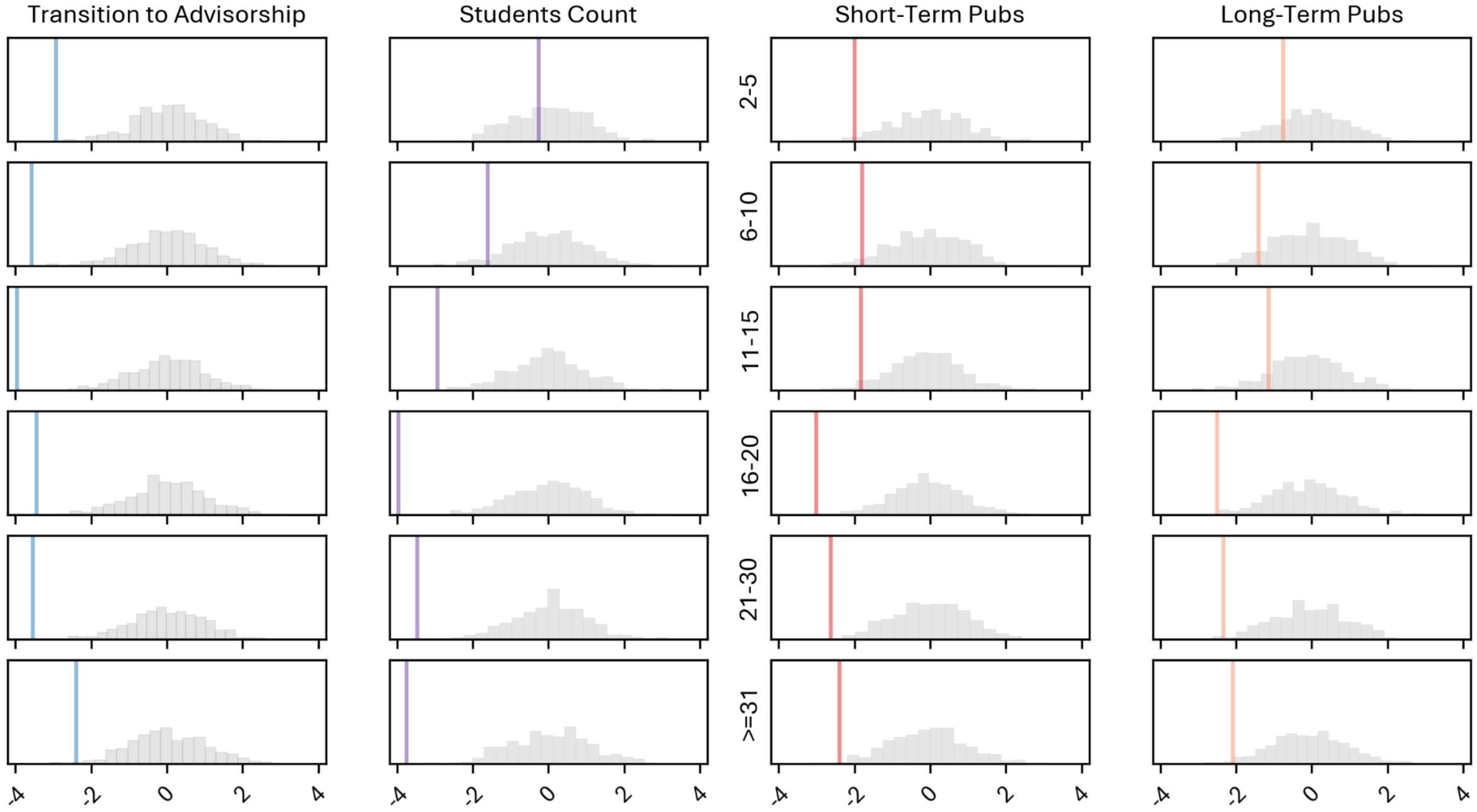

Figure A3. Permutation Test of Intellectual Birth-Order Effects (T-Statistics)

While we cannot conclusively establish causality due to potential unobserved selection and the nature of our question, the convergence of evidence across multiple robustness checks, including propensity score weighting and permutation tests that address different sources of bias, help to partly address the endogeneity problem and lend credibility to a causal interpretation.